\documentclass[twocolumn,english,journal]{IEEEtran}
\usepackage[T1]{fontenc}
\usepackage{color}
\usepackage{babel}
\usepackage{array}
\usepackage{float}
\usepackage{units}
\usepackage{amsmath}
\usepackage{amsthm}
\usepackage{amssymb}
\usepackage{graphicx}
\PassOptionsToPackage{normalem}{ulem}
\usepackage{ulem}
\usepackage[unicode=true,
 bookmarks=false,
 breaklinks=false,pdfborder={0 0 0},pdfborderstyle={},backref=false,colorlinks=false]
 {hyperref}
\hypersetup{pdftitle={Your Title},
 pdfauthor={Your Name},
 pdfpagelayout=OneColumn, pdfnewwindow=true, pdfstartview=XYZ, plainpages=false}

\makeatletter

\providecommand{\tabularnewline}{\\}
\floatstyle{ruled}
\newfloat{algorithm}{tbp}{loa}
\providecommand{\algorithmname}{Algorithm}
\floatname{algorithm}{\protect\algorithmname}

\theoremstyle{plain}
\newtheorem{thm}{\protect\theoremname}
\theoremstyle{plain}
\newtheorem{cor}[thm]{\protect\corollaryname}
\theoremstyle{remark}
\newtheorem{rem}[thm]{\protect\remarkname}
\theoremstyle{definition}
\newtheorem{defn}[thm]{\protect\definitionname}
\theoremstyle{plain}
\newtheorem{lem}[thm]{\protect\lemmaname}

\usepackage{dsfont}
\usepackage{algorithm,algpseudocode}
\usepackage{caption}
\usepackage[caption=false,font=footnotesize]{subfig}
\usepackage{cite}
\usepackage{balance}
\usepackage{graphicx,cite,flushend}
\PassOptionsToPackage{normalem}{ulem}
\usepackage{ulem}
\usepackage{xcolor}
\newcommand\crule[3][black]{\textcolor{#1}{\rule{#2}{#3}}}

\@ifundefined{showcaptionsetup}{}{%
 \PassOptionsToPackage{caption=false}{subfig}}
\usepackage{subfig}
\makeatother

\providecommand{\corollaryname}{Corollary}
\providecommand{\definitionname}{Definition}
\providecommand{\lemmaname}{Lemma}
\providecommand{\remarkname}{Remark}
\providecommand{\theoremname}{Theorem}

\begin{document}

\title{Backpressure on the Backbone: A Lightweight, Non-intrusive Traffic
Engineering Approach }

\author{Christos Liaskos, Xenofontas Dimitropoulos and Leandros Tassiulas
\thanks{C. Liaskos is with the Foundation for Research and Technology Hellas
(FORTH), Institute of Computer Science, N. Plastira 100, Vassilika
Vouton, GR-700 13 Heraklion, Crete, Greece. e-mail: cliaskos@ics.forth.gr.}\thanks{X. Dimitropoulos is with the Computer Science Department, University
of Crete, and the Foundation for Research and Technology Hellas (FORTH),
Institute of Computer Science, N. Plastira 100, Vassilika Vouton,
GR-700 13 Heraklion, Crete, Greece. e-mails: fontas@csd.uoc.gr, fontas@ics.forth.gr.}\thanks{L. Tassiulas is with the School of Engineering and Applied Science,
Yale University, P.O. Box 208263, New Haven, CT 06520, e-mail: leandros.tassiulas@yale.edu.}\thanks{This work was funded by the European Research Council, grant EU338402,
project NetVOLUTION (http://www.netvolution.eu). }}
\maketitle
\begin{abstract}
The present study proposes a novel collaborative traffic engineering
scheme for networks of autonomous systems. Backpressure routing principles
are used for deriving priority routing rules that optimally stabilize
a network, while maximizing its throughput under latency considerations.
The routing rules are deployed to the network following simple SDN
principles. The proposed scheme requires minimal, infrequent interaction
with a central controller, limiting its imposed workload. Furthermore,
it respects the internal structure of the autonomous systems and their
existing peering relations. In addition, it co-exists smoothly with
underlying distance vector-based routing schemes. The proposed scheme
combines simplicity with substantial gains in served transit traffic
volume, as shown by simulations in realistic setups and proven via
mathematical analysis.
\end{abstract}


\IEEEpeerreviewmaketitle{}

\section{Introduction}

\IEEEPARstart{S}{oftware}-Defined Networking (SDN) can imbue the
network management process with an unparalleled level of state monitoring
and control. The ability to migrate the routing elements of a network
from closed, static hardware solutions towards an open, re-programmable
paradigm is expected to promote significantly the adaptivity to time-variant
demand patterns, eventually yielding a healthy and constant innovation
rate. The OpenFlow protocol and assorted hardware~\cite{McKeown.2008},
which enables an administrative authority to centrally monitor a network
and deploy fitting routing strategies, has already produced significant
gains in a wide set of application scenarios~\cite{Hong.2013,Agarwal.2013}.

\textcolor{black}{Traffic engineering (TE) constitutes one of the
networking aspects that has benefited the most from the adoption of
centralized control. The objective of TE is the real-time grooming
of data flows, in order to provide the best possible quality of service
on a given physical infrastructure. Therefore, the success of TE depends
on the ability to obtain accurate snapshots of the network's state
in little time, as well as to react rapidly to state changes~\cite{Kandula.2005}.
Given that SDN offers a unique advantage in terms of centralized network
monitoring and fine-grained actuation, it is not surprising that the
TE and SDN combination has been highly fruitful~\cite{Curtis.2011b}.
Early industrial applications have yielded considerable gains in the
efficient use of network resources, such as link utilization~\cite{Jain.2013}
and application-specific network service~\cite{Hong.2013}. }

\textcolor{black}{Given the success of SDN solutions in proprietary
networks such as datacenters~\cite{Jain.2013,Hong.2013}, research
aims to bring its benefits to the Internet routing~\cite{D4D}. Current
studies promise the provision of end-to-end QoS guarantees at Internet-level~\cite{kotronisCXP,SDX.2015},
fast and secure convergence of global TE~\cite{PavlosMAMA2016},
and enabling the continuous evolution of Internet routing~\cite{KontronisNetvolution}.
Nonetheless, adopting such solutions at global level faces the challenge
of multi-domain coordination. The Internet comprises Autonomous Systems
(ASes), which in their majority are privately held and managed organizations.
On one hand, each AS may have its own hardware, internal network control
system, as well as financial status and business model. On the other
hand, related solutions promise impressive performance but typically
require changes in the AS internal state of affairs, at least partially~\cite{PavlosMAMA2016,kotronisCXP,SDX.2015}.
Thus, while collaborative TE may be to the financial interest of several
ASes~\cite{Castro.2014,Gyarmati.}, more intermediate steps may be
required for promoting it.}

\textcolor{black}{The present study contributes a light collaborative
TE solution for AS collaboration, where participants maintain full
control of their own networks. The methodology consists of applying
the principles of Backpressure (BPR) routing to a network of ASes~\cite{Tassiulas.1992}.
BPR is well-known for its simple principles, leading to analytically-proven
throughput-optimality nonetheless. Within the proposed system, collaborating
ASes inform a central service of internal congestion events that they
experience. This information can be derived in any manner the AS presently
supports. Moreover, it refers to the destination of the congested
traffic only and, therefore, is partial and of temporal validity.
In return, congested ASes receive a small set of proposed priority
routing rules derived by BPR. The core BPR principle is the offloading
of the congested traffic to the least loaded neighboring AS. ASes
can decide upon the application of the proposed rules and proceed
to their deployment in any custom manner. No further details on the
inner workings of an AS need to be shared. The proposed scheme can
respect existing routing algorithms, peering preferences and latency
considerations, while offering analytically-proven throughput maximization
and network stabilization potential. Moreover, throughput maximization
can translate to substantially increased transit traffic, which can
be of financial interest to the collaborating ASes. The proposed scheme
is intended as an intermediate step towards fostering AS collaboration.
Its aim is to serve as a stepping stone towards more advanced SDN
schemes for AS collaboration~\cite{SDX.2015,kotronisCXP,KontronisNetvolution}.
It prioritizes compatibility with existing AS equipment and lays a
foundation for centralized, SDN-based inter-AS routing orchestration.}

\textcolor{black}{The remainder of this paper is as follows. Prerequisites
are given in Section \ref{sec:Prerequisites}. The system model and
the analysis follow in Sections \ref{sec:The-Proposed-System}~and~\ref{sec:Analysis}.
Evaluation via simulations takes place in Section \ref{sec:Evaluation}.
Section \ref{sec:Related-Work} presents the related studies and the
paper is concluded in Section \ref{sec:Conclusion-and-Future}. \vspace{-5bp}
}

\section{\textcolor{black}{Prerequisites\label{sec:Prerequisites}\vspace{-5bp}
}}

\textcolor{black}{}
\begin{table}[t]
\begin{centering}
\textcolor{black}{\caption{Summary of Notation\label{tab:Summary-of-Notation}}
}
\par\end{centering}
\centering{}\textcolor{black}{}%
\begin{tabular*}{1\columnwidth}{@{\extracolsep{\fill}}|>{\centering}p{1.5cm}|>{\centering}p{6.5cm}|}
\hline
\textcolor{black}{Symbol} & \textcolor{black}{Explanation}\tabularnewline
\hline
\hline
\textcolor{black}{$U_{(n,c)}(t)$} & \textcolor{black}{The aggregate traffic accumulated within a network
node $n$ at time $t$, destined towards node $c$.}\tabularnewline
\hline
\textcolor{black}{$T$} & \textcolor{black}{The network state monitoring and actuation period.}\tabularnewline
\hline
\textcolor{black}{$O_{(n,c)}^{t\to t+T}$} & \textcolor{black}{Data volume outgoing from node $n$ to $c$ in the
interval $\left[t,\,t+T\right]$.}\tabularnewline
\hline
\textcolor{black}{$I_{(n,c)}^{t\to t+T}$} & \textcolor{black}{Data volume incoming to node $n$, destined to $c$,
in the interval $\left[t,\,t+T\right]$.}\tabularnewline
\hline
\textcolor{black}{$G_{(n,c)}^{t\to t+T}$} & \textcolor{black}{Data volume generated at node $n$, (or arriving
to $n$ from a network-external source), destined to $c$, in the
interval $\left[t,\,t+T\right]$.}\tabularnewline
\hline
\textcolor{black}{$\mu_{l}^{(c)}(t)$} & \textcolor{black}{The maximum allowed bitrate at time $t$ over a
network link $l$, carrying traffic destined to node $c$. For wired
backbone links, $\mu_{l}^{(c)}$ is the nominal capacity of $l$,
and is invariant of $t$.}\tabularnewline
\hline
\textcolor{black}{$\overline{\lambda_{(n,c)}}(t)$} & \textcolor{black}{The average traffic production rate at node $n$
destined to $c$ for the time interval $[t,t+T]$.}\tabularnewline
\hline
\textcolor{black}{$n'\gets\mathcal{P}(n)$} & \textcolor{black}{The application of a routing policy $\mathcal{P}$
on data residing at node $n$, returning the next intended hop $n'$.}\tabularnewline
\hline
\textcolor{black}{$\overrightarrow{T}\left\{ \mathcal{P}^{\circ m}(n)\right\} $} & \textcolor{black}{The set of traversed nodes when applying $m$ times
the policy $\mathcal{P}$ on node $n$.}\tabularnewline
\hline
\end{tabular*}
\end{table}
\textcolor{black}{}
\begin{algorithm}[t]
\textcolor{black}{\footnotesize{}\begin{algorithmic}[1]
\Procedure{SBPR}{$network\_state(t=mT|m\in\mathbf{N})$}
\For {each node $c$}
\Comment Assign traffic type to links.
\For {each link $l$}
\State $c_l^{*}(t)\gets argmax_{c}\{U_{source(l),c}-U_{dest(l),c}\}$
\State $\Delta Q^{*}_l(t)\gets max_{c}\{0,U_{source(l),c}-U_{dest(l),c}\}$
\EndFor
\EndFor \Comment Calculate optimal transfer rates per link.
\State $\mathbf{\mu}^{*}(t)\gets argmax_{\mathbf{\mu}}\sum_{\forall l}\mu_{l}(t)\cdot \Delta Q^{*}_{l}(t)$
\For {each link $l:\Delta Q^{*}_l(t)>0$}
\State Transfer data destined to $c_{l}^{*}(t)$ with rate $\mu_{l}^{*}(t)$.
\EndFor
\EndProcedure
\end{algorithmic}}{\footnotesize \par}

\textcolor{black}{\footnotesize{}\caption{\label{alg:classicBP}The Standard Backpressure Routing algorithm.}
}{\footnotesize \par}
\end{algorithm}

\textcolor{black}{An important term in networking studies is the notion
of }\textcolor{black}{\emph{network stability}}\textcolor{black}{.
It is defined as the ability of a routing policy to keep all network
queues bounded, provided that the input load is within the network's
traffic dispatch ability, i.e. within its }\textcolor{black}{\emph{stability
region}}\textcolor{black}{. Let $U_{(n,c)}(t)$ denote the aggregate
traffic accumulated within a network node $n$ at time $t$, destined
towards node $c$. Stability is then defined as~\cite[p. 24]{Georgiadis.2005}:
\begin{equation}
\underset{\tau\to+\infty}{lim\,sup}\frac{1}{\tau}\sum_{t=1}^{\tau}E\left\{ U_{(n,c)}(t)\right\} <\infty,\,\forall n,c\label{eq:stability-defintion}
\end{equation}
where $\tau$ is the time horizon and $E\left\{ *\right\} $ denotes
averaging over any probabilistic factors present in the system. }

\textcolor{black}{The Standard BPR routing algorithm (SBPR, Algorithm~\ref{alg:classicBP})
has been proven to optimally stabilize a network, i.e., keeping it
within its stability region and maximizing its throughput~\cite{Neely.2003c,Neely.2005b}.
Its operating principle is simple: a node $n$ offloads traffic towards
a destination $c$ by redirecting it to its less congested, immediate
neighbor (lines $2-7$). If the interconnecting links have non-constant
bandwidth (i.e., wireless links), a selection of valid rates takes
place in line $8$, and the data is transferred in lines $9-11$.}

\textcolor{black}{The SBPR algorithm was derived analytically, using
a well-known network stability framework, i.e., the Lyapunov Drift.
This approach defines a quadratic function of the form:
\begin{equation}
L(t)=\underset{n}{\sum}\underset{c}{\sum}U_{(n,c)}^{2}(t)
\end{equation}
The goal is then to deduce the bounds of $\Delta L(t)=E\left\{ L(t+T)-L(t)\right\} ,$
which describes the evolution of the network queue levels over a time
period $T$. The }\textcolor{black}{\emph{Lyapunov stability theorem}}\textcolor{black}{{}
states that if it holds \cite[p. 50]{Georgiadis.2005}:
\begin{equation}
\Delta L(t)\le B-\epsilon\cdot\underset{n}{\sum}\underset{c}{\sum}U_{(n,c)}(t)\label{eq:LyastabilityCriteria}
\end{equation}
for any two positive quantities, $B$ and $\epsilon$, then the network
is stable and average queue size of inequality~(\ref{eq:stability-defintion})
is bounded by $\nicefrac{B}{\epsilon}$ instead of drifting towards
infinity.}

\textcolor{black}{The }\textcolor{black}{\emph{BPR }}\textcolor{black}{class
of algorithms defines a routing policy that complies with the stability
criteria of inequality~(\ref{eq:LyastabilityCriteria}). Its goal
is to minimize the lower bound of $\Delta L(t)\,\forall t$, effectively
suppressing the average queue level within the network. The analytical
approach, followed by related studies~\cite{Leonardi.,McKeown.1999},
is outlined as follows. }

\textcolor{black}{Firstly, the queue dynamics are expressed by the
following relation, $\forall n,c$~\cite[p. 54]{Georgiadis.2005}:
\begin{equation}
\underset{_{(n,c)}}{U}^{(t+T)}=max\left\{ 0,U_{(n,c)}^{(t)}-O_{(n,c)}^{t\to t+T}\right\} +I{}_{(n,c)}^{t\to t+T}+G{}_{(n,c)}^{t\to t+T}\label{eq:strictQdynamics}
\end{equation}
where $O_{(n,c)}^{t\to t+T}$, $I_{(n,c)}^{t\to t+T}$ and $G_{(n,c)}^{t\to t+T}$
denotes outgoing, incoming and locally generated data, as summarized
in Table~\ref{tab:Summary-of-Notation}. Notice that these quantities
refer to any traffic exchange process in general, and not just to
traffic exchanged due to BPR routing.}

\textcolor{black}{Secondly, both sides of equation~(\ref{eq:strictQdynamics})
are squared, and a series of relaxations are applied on its right
hand side (RHS). These relaxations are based on the identity~\cite[p. 53]{Georgiadis.2005}:
\begin{equation}
V\le max\left\{ 0,\,U-\mu\right\} +A\Rightarrow V^{2}\le U^{2}+\mu^{2}+A^{2}-2U\cdot(\mu-A)\label{eq:identity}
\end{equation}
as well as on the following inequalities, $\forall n,c$:
\begin{equation}
O_{(n,c)}^{t\to t+T}\le\underset{l:\,source(l)=n}{\sum}\int_{t}^{t+T}\mu_{l}^{(c)}(t)\cdot dt\label{eq:relax1}
\end{equation}
\begin{equation}
I_{(n,c)}^{t\to t+T}\le\underset{l:\,destination(l)=n}{\sum}\int_{t}^{t+T}\mu_{l}^{(c)}(t)\cdot dt\label{eq:relax2}
\end{equation}
where $\mu_{l}^{(c)}(t)$ is the available capacity of a network link
$l$ at time $t$, that can carry traffic destined to node $c$. }

\textcolor{black}{Thus, one derives an inequality of the form of relation~(\ref{eq:LyastabilityCriteria}).
Further relaxation by substituting all $\mu_{l}^{(c)}$ and $G_{(n,c)}^{t\to t+T}$
with maximum, nominal values yields compliance with the Lyapunov stability
theorem. Furthermore, it is deduced that the upper bound of relation~(\ref{eq:LyastabilityCriteria})
can be minimized when maximizing the quantity~\cite[p. 66]{Georgiadis.2005}:
\begin{equation}
\underset{\forall n,k,c}{\sum}\mu_{l:source(n)\to dest(k)}^{(c)}(t)\cdot\left(U_{(n,c)}(t)-U_{(k,c)}(t)\right)\label{eq:optPursuitBPR}
\end{equation}
 The SBPR algorithm (Algorithm~\ref{alg:classicBP}) essentially
expresses the optimization pursuit of relation~(\ref{eq:optPursuitBPR})
at lines $2-7$. }

\textcolor{black}{The BPR class of algorithms has found extensive
use in packet switching hardware, wireless ad hoc networks and satellite
systems due to their throughput optimality trait~\cite{Leonardi.,McKeown.1999,Neely.2005b}.
However, if used as a standalone routing policy with no other support,
BPR can yield increased latency. This is due to the fact that queue
sizes must build up in order to obtain backpressure potential, $\Delta U$,
yielding increased queuing delay. On the other hand, when the queue
sizes are low, routing may resemble a random walk, accentuating propagation
delays. Furthermore, if $T$ is smaller than the full trip time of
a packet, loops may appear in its path. Newer algorithms of the BPR
class mitigate these issues and take into account latency considerations.
For example, authors in~\cite{Ying.2011b} restrict steps $3-5$
of Algorithm~\ref{alg:classicBP} only within a subset of links that
offer a bounded maximum number of hops towards the target. Other studies
have shown that simply altering the queuing discipline from FIFO to
LIFO yields considerable latency gains~\cite{Huang.2013b,LongboHuang.2011,Moeller.}.
Finally, it is worth noting that the BPR class can be made TCP compatible
in a straightforward fashion~\cite{Seferoglu.}.}

\textcolor{black}{Another major advantage of BPR is the fair distribution
of traffic over the nodes. Particularly, it has been shown that the
closer a node is to the gateway(s), the more the transit traffic will
pass through it over the time interval $T$~\cite{Moeller.}. Thus,
nodes surrounding a gateway will naturally receive more transit traffic
than a distant node. Nonetheless: i) nodes with similar distance and
connectivity to the gateway(s) will receive similar transit traffic
volume, while ii) all nodes in general will receive surplus traffic~\cite{Moeller.}.
In addition, the variance of the transit traffic served by each node
is minimal and very close to their long-term average~\cite{LongboHuang.2011}.
Therefore, BPR shares the transit traffic fairly among the network
nodes, with regard to their connectivity and their placement within
the topology.}

\section{\textcolor{black}{The System Model\label{sec:The-Proposed-System}}}

\textcolor{black}{}
\begin{figure}[t]
\begin{centering}
\textcolor{black}{\includegraphics[width=1\columnwidth]{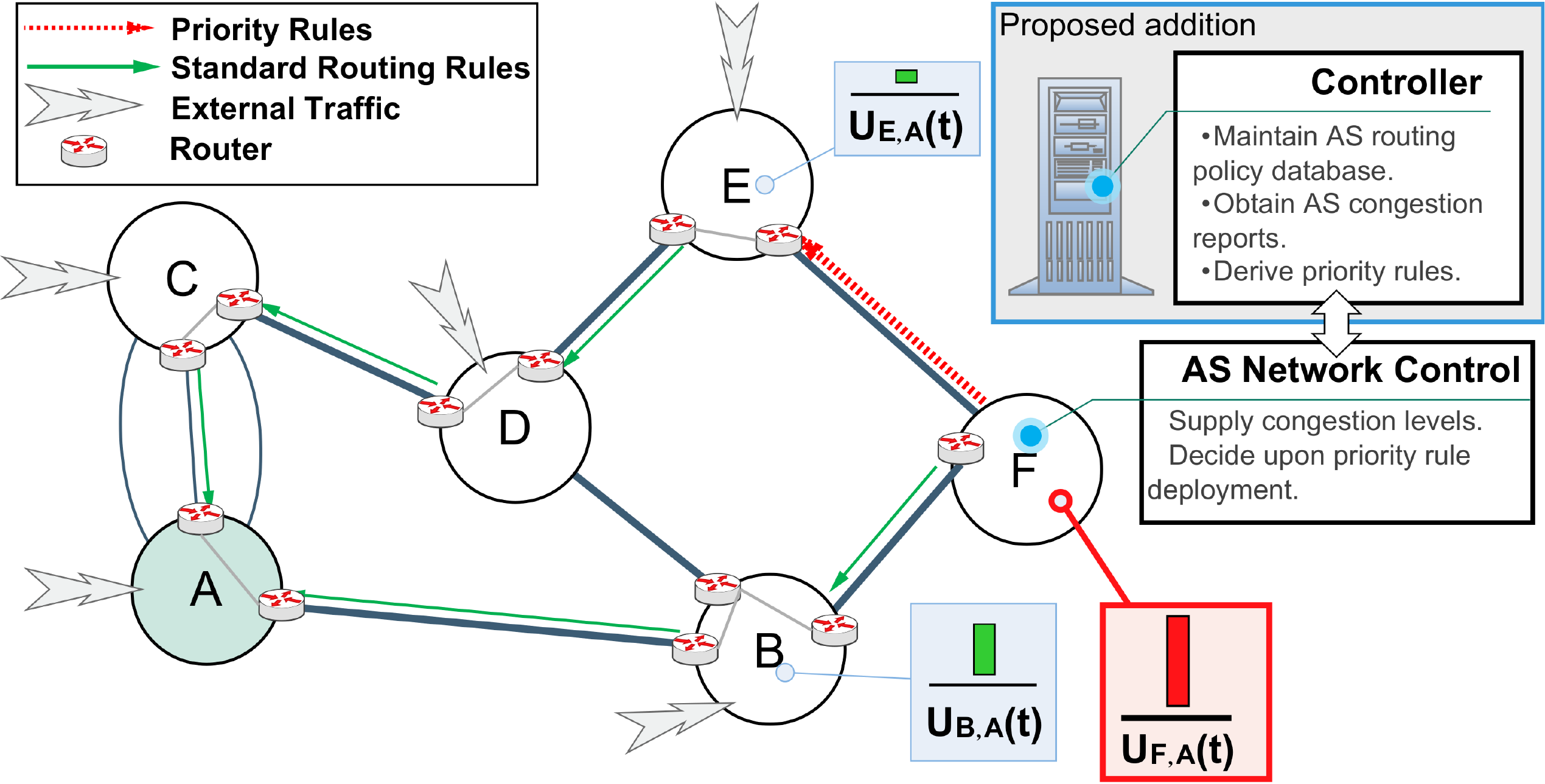}}
\par\end{centering}
\textcolor{black}{\caption{\textcolor{black}{\label{fig:Setup}The employed system setup. A network
of ASes, A-F, uses BPR-derived routing rules on top of its standard
routing scheme. A centralized control entity supplies the temporary
priority rules on demand, while existing AS Network Control Systems
retain their monitoring and routing jurisdiction.}}
}
\end{figure}

\textcolor{black}{The present paper proposes the use of BPR policies
in backbone networks. The assumed setup, given in Fig.~\ref{fig:Setup},
considers a network comprising border routers organized in ASes (}\textcolor{black}{\uline{nodes}}\textcolor{black}{).
These ASes constitute a cluster of collaborating, same-tier, transit
service providers (ISPs). The ASes exchange traffic with the rest
of the Internet normally (i.e., they are connected to other, non-collaborating
ASes) via arbitrary links. The proposed scheme seeks to promote the
creation of such collaboration clusters, by yielding a significant
increase in the total transit traffic serviced by the cluster. }

\textcolor{black}{The proposed scheme prioritizes minimal operational
and equipment changes in the AS cluster. Towards this end: i) each
AS is allowed to retain its existing Network Control System (NCS),
which comprises AS-local router-level topology maintenance, congestion
measurement services, link quality monitoring and routing rule management.
The proposed scheme essentially operates on top of any set of NCSes
(legacy or SDN-based, such as ONOS~\cite{onos} or Beehive \cite{beehive}),
making use of their existing functionality. ii) The cluster of ASes
may have any load-invariant routing policy, e.g., Distance Vector
Routing (DVR) enforced via the Border Gateway Protocol (i/eBGP)~\cite{Akashi.2008}.
The proposed scheme assumes no changes or obstruction to the BGP operation.
ASes within the collaboration cluster exchange BGP updates with other
ASes normally, whether these are members of the collaboration cluster
or not. Routing policies (i.e., AS routing preferences) are allowed,
as per the existing status quo in inter-AS routing. iii) The proposed
scheme functions by installing priority routing rules at }\textcolor{black}{\emph{adjacent}}\textcolor{black}{{}
border routers, essentially operating on top of the underlying BGP
routing. These priority rules are of }\textcolor{black}{\emph{temporal}}\textcolor{black}{{}
value and aim at alleviating AS-local increases in traffic load. As
such, the priority rules }\textcolor{black}{\emph{do not}}\textcolor{black}{{}
propagate in the form of updates, either within or outside the cluster. }

\textcolor{black}{The proposed scheme considers the addition of a
central }\textcolor{black}{\uline{Controller}}\textcolor{black}{,
which supplies each AS-local NCS with the proposed routing rules upon
demand. An NCS is assumed to monitor its internal congestion level
using its existing components. On the event of increased congestion
(or periodically), it forwards the congestion measurement to the Controller,
obtaining a set of proposed priority routing rules in return. The
NCS then decides upon their deployment or rejection. Exemplary reasons
for rejection are a sudden change in the AS routing policies or reference
to faulty/flapping links, as detected by the NCS. Subsequently, the
Controller is updated on the future AS policy and set of usable links. }

\textcolor{black}{The traffic volume metric monitored by the NCS of
each AS, $n$, is essentially $U_{(n,c)}(t)$ (cf. Table~\ref{tab:Summary-of-Notation}).
Therefore, we notice that the congestion information passed to the
Controller is }\textcolor{black}{\uline{partial}}\textcolor{black}{{}
and specific to the congestion event. Obtaining $U_{(n,c)}(t)$ is
handled by an NCS, using any existing module or approach. For instance,
the }\textcolor{black}{\emph{router-state polling}}\textcolor{black}{{}
technique measures the ingress and egress traffic volume that has
traversed the interfaces of the AS border routers over a period of
time. The quantity $U_{(n,c)}(t)$ is then calculated simply subtracting
the total egress from the total ingress load, for every AS $c\ne n$
in the cluster. This type of monitoring can be accomplished in a very
short amount of time ($1-5$ sec), even for very large networks~\cite{Tootoonchian.2010}.
Without loss of generality, we will assume that the Controller is
supplied with the $U_{(n,c)}(t)$ measurements with period $T$, for
every $n$ and any destination $c$ that exhibits congested traffic
within $n$. }

\textcolor{black}{The Controller also acts as a repository holding
the inter-AS routing preferences within the collaboration cluster.
Each collaborating AS is assumed to regularly submit and update its
routing preferences to this repository. The submitted information
remains private and is used only internally by the Controller to derive
proper priority rules. The collaborating ASes may also inform the
controller of their topology, i.e., their inter-AS }\textcolor{black}{\emph{usable}}\textcolor{black}{{}
links and corresponding capacities. Notice this information can be
presently approximated from public databases as well~\cite{CAIDAIDTK.2016}.
Having the routing preferences, the topology and the $U_{(n,c)}(t)$
information at its disposal, the Controller executes a BPR algorithm
(e.g., Algorithm~\ref{alg:classicBP}) at AS-level. }

\textcolor{black}{An example is shown in Fig.~\ref{fig:Setup}, where
the Controller seeks to offload the traffic volume $U_{(F,A)}(t)$
at node $F$. A BPR algorithm is executed, which deduces that traffic
from $F$ towards $A$ should better be offloaded to neighboring node
$E$ for the time being. A corresponding priority routing proposal
is passed to AS $F$, which handles its deployment. The installed
priority rule takes precedence over all other routing rules pertaining
to link $l_{FE}$. Other routers within AS $F$ (e.g., using iBGP)
are then instructed to forward traffic for AS $A$ towards the selected
border router(s) by the active NCS. }

\textcolor{black}{Peering agreements and routing preferences are handled
during the execution of the BPR algorithm. For example, returning
to the example of Fig. \ref{fig:Setup}, the controller would not
propose the illustrated flow rule if it was disallowed by the respective
peering preferences/agreements affecting $F$ and $E$. In other words,
when the BPR-variant searches for neighbors $s\in S:\{U_{(s,A)}(t)<U_{(F,A)}(t)\}$,
the search is assumed to be limited to nodes that comply with any
form of preference of agreement. }

\textcolor{black}{The management of the derived priority rules follows
the general principles of the OpenFlow protocol~\cite{McKeown.2008}.
Their lifetime is ended by the AS when the local congestion drops
beyond a safety-level, or if the rule has remained idle for a period
of time. Thus, the AS system reverts to its DVR routing policy as
soon as possible. The Controller participates in the management of
the rules as well. Similarly to the OpenFlow case, basic concerns
when installing the priority routing rules are: i) the avoidance of
routing loops, and ii) the protection against partial or asynchronous
rule deployment. In this aspect, the Controller can implement existing
algorithms for the priority rule derivation, which promise scalability
and disruption-free network operation~\cite{vanbever2011seamless}.
An interesting trait of the proposed scheme is that it can guarantee
a basic, loop-free operation, even without such a rule management
mechanism. This trait is based on a smart filtering of AS-neighbors
when applying BPR and is analyzed in Section~\ref{sec:Analysis}.}

\textcolor{black}{It is also worth noting the motives for participation
of an AS to the proposed collaboration scheme, which derive from the
inherent properties of BPR:}
\begin{itemize}
\item \textcolor{black}{Due to throughput optimality, the nodes are expected
to handle more transit traffic, i.e., incoming to the cluster of collaborating
nodes, using the same links, postponing the need for capacity upgrades.}
\item \textcolor{black}{The incoming transit traffic is naturally shared
fairly among the collaborating nodes, considering their connectivity
and position in the topology, as discussed in Section~\ref{sec:Prerequisites}.
This can be an attractive trait in the cases where the monetary profit
of the ASes is strongly related to the served traffic~\cite{Gyarmati.}.
In such cases, the pricing and profit model can remain unaltered,
without necessarily requiring a profit-sharing mechanism.}
\item \textcolor{black}{Natural traffic sharing and throughput optimality
may also imply increased resilience against link-outages, attributed
to benign causes (e.g., link failures, BGP session resets, link flooding
due to flash-crowds~\cite{Thapngam.}) or attacks (e.g.,~\cite{Kang.2013,Studer.2009}).
The latter have recently shown potential to even cut-off ASes from
the Internet~\cite{.q}. In this case, collaboration can potentially
postpone the need for sophisticated-yet very intrusive-defense mechanisms~\cite{Lee.2013}.}
\item \textcolor{black}{The nodes remain autonomous, and need not reveal
their internal structure/equipment/client set or relinquish control
over it.}
\item \textcolor{black}{BPR relies on node-local traffic state information
only (cf. Algorithm~\ref{alg:classicBP}). The required information
is a common congestion metric. }
\item \textcolor{black}{Controller failures are not critical, as on such
events the ASes continue their normal, BGP-based operation.}
\end{itemize}
\textcolor{black}{In summary, the key-benefit of the proposed scheme
is that it promises increased stability and transit traffic, while
incurring minimal changes in the operation of the ASes. The proposed
scheme follows some general OpenFlow operation principles, namely
the central derivation of priority flow rules, but does not require
upgrades to OpenFlow equipment. To the contrary, it is designed to
work with the existing NCS that an AS has adopted. The main implementation
cost for the proposed scheme is thus that of the Controller. In this
aspect, the operation that the Controller should support does not
differ significantly from a web service, which constitutes an indication
of limited capital and operational expenditure.}

\section{\textcolor{black}{Analysis\label{sec:Analysis}}}

\textcolor{black}{The target of the analysis is to evaluate the use
of BPR routing in backbone networks. To this end, the following sub-objectives
are studied separately:}
\begin{itemize}
\item \textcolor{black}{The potential of BPR algorithms to take future traffic
states into account.}
\item \textcolor{black}{The system's dependence on the controller. To this
end, we study the effects of the monitoring and actuation period $T$
on the network drift bound.}
\item \textcolor{black}{The loop-free, latency-aware cooperation between
BPR and the underlying routing scheme, considering asynchronous or
partial deployments of BPR rules. }
\end{itemize}
\textcolor{black}{Finally, we allow for at most a single priority
flow rule per physical network link, which will be shown to be sufficient
for significant performance gains. }

\subsection{\textcolor{black}{On traffic forecast-awareness in BPR routing.}}

\textcolor{black}{SBPR can serve as the BPR algorithm in the proposed
system model of Section~\ref{sec:The-Proposed-System}. SBPR operates
without knowledge of future traffic patterns, based on current congestion
states only. Nonetheless, backbone links are characterized by high
capacity. At such nominal data rates, the congestion distribution
of ASes could vary rapidly. Moreover, ISPs already employ Internet
traffic forecasts to their present TE approaches~\cite{Otoshi.2015,Steland.2015}.
These forecasts are quite accurate, indicatively yielding $1-3$~\%
error for $T=1\to5$ min and $3-5$~\% for $T=1$ hour~\cite{Cortez.2010}.
Thus, the posed question is whether a new BPR algorithm can consider
this additional information, without altering the properties of SBPR.}

\textcolor{black}{We begin the analysis by simplifying the RHS of
relation~(\ref{eq:relax2}), based on the fact that the network links
have time-invariant bandwidth:
\begin{equation}
I_{(n,c)}^{t\to t+T}\le\underset{l:\,d(l)=n}{\sum}\int_{t}^{t+T}\mu_{l}^{(c)}dt=T\cdot\underset{l:\,d(l)=n}{\sum}\mu_{l}^{(c)}\label{eq:relax1-1}
\end{equation}
In the same manner, the RHS of relation~(\ref{eq:relax1}) becomes:
\begin{equation}
O_{(n,c)}^{t\to t+T}\le\underset{l:\,s(l)=n}{\sum}\int_{t}^{t+T}\mu_{l}^{(c)}dt=T\cdot\underset{l:\,s(l)=n}{\sum}\mu_{l_{nb(n)}}^{(c)}\label{eq:relax2-1}
\end{equation}
where $b(n)$ represents the neighboring node of $n$, at the end
of each outgoing link. The considered links are the ones compliant
with any bilateral routing preferences. Moreover, for ease of presentation
we set:}

\textcolor{black}{
\begin{equation}
\mathring{\mu}_{(n)}^{(c)}=\underset{l:\,s(l)=n}{\sum}\mu_{l_{nb(n)}}^{(c)}
\end{equation}
We proceed to apply identity~(\ref{eq:identity}) to eq.~(\ref{eq:strictQdynamics}):
\begin{multline}
\hspace{-5bp}\hspace{-8bp}U_{(n,c)}^{2}(t+T)\le U_{(n,c)}^{2}(t)+\left[O_{(n,c)}^{t\to t+T}\right]^{2}+\left[I_{(n,c)}^{t\to t+T}+G{}_{(n,c)}^{t\to t+T}\right]^{2}\\
-2\cdot U_{(n,c)}(t)\cdot\left[O_{(n,c)}^{t\to t+T}-I{}_{(n,c)}^{t\to t+T}-G{}_{(n,c)}^{t\to t+T}\right]
\end{multline}
Using the updated relaxations (\ref{eq:relax1-1}) and (\ref{eq:relax2-1})
and setting $\Delta U_{(n,c)}^{2}(t)=U_{(n,c)}^{2}(t+T)-U_{(n,c)}^{2}(t)$
for brevity:~~~
\begin{multline}
\hspace{-5bp}\Delta U_{(n,c)}^{2}(t)\le T^{2}\cdot\left({\color{blue}{\color{black}\mathring{\mu}_{(n)}^{(c)}}}\right)^{2}+\left[T\cdot\underset{l:\,d(l)=n}{\sum}\mu_{l}^{(c)}+G{}_{(n,c)}^{t\to t+T}\right]^{2}\\
-2\cdot U_{(n,c)}(t)\cdot\left[T\cdot{\color{black}{\color{blue}\mathring{{\color{black}\mu}}}_{{\color{blue}{\color{black}(n)}}}^{{\color{blue}{\color{black}(c)}}}}-T\cdot\underset{l:\,d(l)=n}{\sum}\mu_{l}^{(c)}-G{}_{(n,c)}^{t\to t+T}\right]\label{eq:RHS_raw}
\end{multline}
It is not difficult to show that the RHS of inequality (\ref{eq:RHS_raw})
can be reorganized as:
\begin{multline}
\Delta U_{(n,c)}^{2}(t)\le\left[T\cdot\underset{l:\,d(l)=n}{\sum}\mu_{l}^{(c)}+U_{(n,c)}(t)+G{}_{(n,c)}^{t\to t+T}\right]^{2}\\
+\left[T\cdot{\color{black}{\color{blue}\mathring{{\color{black}\mu}}}_{{\color{blue}{\color{black}(n)}}}^{{\color{blue}{\color{black}(c)}}}}-U_{(n,c)}(t)\right]^{2}-2\cdot U_{(n,c)}^{2}(t)
\end{multline}
Summing both sides $\forall n,c$ and reminding that $\Delta L(t)=\underset{\forall n}{\sum}\underset{\forall c}{\sum}\Delta U_{(n,c)}^{2}(t)$:
\begin{multline}
\Delta L(t)\le\underset{\forall n}{\sum}\underset{\forall c}{\sum}\underset{(B)}{\left[T\cdot\underset{l:\,d(l)=n}{\sum}\mu_{l}^{(c)}+U_{(n,c)}(t)+G{}_{(n,c)}^{t\to t+T}\right]^{2}}\\
+\underset{\forall n}{\sum}\underset{\forall c}{\sum}\underset{(A)}{\left[T\cdot{\color{black}{\color{blue}\mathring{{\color{black}\mu}}}_{{\color{blue}{\color{black}(n)}}}^{{\color{blue}{\color{black}(c)}}}}-U_{(n,c)}(t)\right]^{2}}-2\cdot\underset{\forall n}{\sum}\underset{\forall c}{\sum}U_{(n,c)}^{2}(t)\label{eq:earlyInsights}
\end{multline}
Relation (\ref{eq:earlyInsights}) provides some early insights on
the minimization of the upper bound of the Lyapunov drift, $\Delta L(t)$.
First of all, the terms (A) and (B) comprise sums of squares, which
can be minimized by ideally nullifying each squared quantity. Term~(A)
advocates for long-lived routing decisions and against bandwidth waste.
Furthermore, outflows from a term (A) will be added to a term within
(B). While the present load of the recipient node, $U_{(n^{*},c)}(t)$
could be low or null, it may be expected to, e.g., generate a considerable
load locally in the time interval $t\to t+T$ (expressed via $G_{(n,c)}^{t\to t+T}$).
To quantify this concerns, we treat the RHS of relation (\ref{eq:earlyInsights})
as a function of the BPR-derived routing decisions ${\color{black}{\color{blue}\mathring{{\color{black}\mu}}}_{{\color{blue}{\color{black}(n)}}}^{{\color{blue}{\color{black}(c)}}}}$
and attempt a straightforward optimization. The ${\color{black}{\color{blue}\mathring{{\color{black}\mu}}}_{{\color{blue}{\color{black}(n)}}}^{{\color{blue}{\color{black}(c)}}}}$
can be initially treated as continuous variables. Once optimal values
have been derived, they can be mapped to the closest of the actually
available options within the network topology. The sufficient conditions
for the presence of a minimum are:
\begin{equation}
\left\{ \begin{array}{cc}
\frac{\partial RHS_{(\ref{eq:earlyInsights})}}{\partial{\color{blue}\mathring{{\color{black}\mu}}_{{\color{black}(n)}}^{{\color{black}(c)}}}}=0 & (i)\\
0<\mathcal{H}\left(\frac{\partial^{2}RHS_{(\ref{eq:earlyInsights})}}{\partial{\color{blue}\mathring{{\color{black}\mu}}_{{\color{black}(n)}}^{{\color{black}(c)}}}\cdot\partial{\color{blue}\mathring{{\color{black}\mu}}_{{\color{black}(k)}}^{{\color{black}(c)}}}}\right)<\infty & (ii)\\
RHS_{(\ref{eq:earlyInsights})}\,\mbox{is convex w.r.t.}\,{\color{blue}\mathring{{\color{black}\mu}}_{{\color{black}(n)}}^{{\color{black}(c)}}}\mbox{,\,}{\color{blue}\mathring{{\color{black}\mu}}_{{\color{black}(k)}}^{{\color{black}(c)}}} & (iii)
\end{array}\right.\label{eq:hessian}
\end{equation}
where $k$ denotes a node, $\mathcal{H}$ is the Hessian matrix~\cite{HiriartUrruty.1984}
and the requirement $0<\mathcal{H}<\infty$ refers to each of its
elements. From condition~(\ref{eq:hessian}-i) we obtain:
\begin{multline}
2T\cdot\left[T\cdot\underset{l:\,d(l)=b(n)}{\sum}\mu_{l}^{(c)}+U_{(b(n),c)}(t)+G{}_{(b(n),c)}^{t\to t+T}\right]\\
+2\cdot T\left[T\cdot{\color{blue}\mathring{{\color{black}\mu}}_{{\color{black}(n)}}^{{\color{black}(c)}}}-U_{(n,c)}(t)\right]=0\Longleftrightarrow
\end{multline}
\begin{multline}
T\cdot\left[\underset{l:\,d(l)=b(n)}{\sum}\mu_{l}^{(c)}+{\color{blue}\mathring{{\color{black}\mu}}_{{\color{black}(n)}}^{{\color{black}(c)}}}\right]\\
-\left[U_{(n,c)}(t)-\left(U_{(b(n),c)}(t)+G{}_{(b(n),c)}^{t\to t+T}\right)\right]=0,\,\forall n,c\label{eq:optimization_goal}
\end{multline}
It is not difficult to show that conditions (\ref{eq:hessian}-ii,
iii) are satisfied, since it holds:
\begin{equation}
\frac{\partial^{2}RHS_{(\ref{eq:earlyInsights})}}{\partial{\color{blue}\mathring{{\color{black}\mu}}_{{\color{black}(n)}}^{{\color{black}(c)}}}\cdot\partial{\color{blue}\mathring{{\color{black}\mu}}_{{\color{black}(k)}}^{{\color{black}(c)}}}}\propto T^{2}>0,\,\forall n,k\label{eq:convex}
\end{equation}
Equation (\ref{eq:optimization_goal}) represents a generalization
over the SBPR (Algorithm~\ref{alg:classicBP}). At first, eq.~(\ref{eq:optimization_goal})
defines a linear system with discrete variables ${\color{blue}\mathring{{\color{black}\mu}}_{{\color{black}(n)}}^{{\color{black}(c)}}}$
and can be solved as such. However, interesting approximations can
be derived, which also exhibit a dependence of the optimal solution
from the network traffic forecasts. }

\textcolor{black}{Firstly, we notice that the term $T\cdot\left[\underset{l:\,d(l)=b(n)}{\sum}\mu_{l}^{(c)}+{\color{blue}\mathring{{\color{black}\mu}}_{{\color{black}(n)}}^{{\color{black}(c)}}}\right]$
includes }\textcolor{black}{\emph{nominal}}\textcolor{black}{{} link
capacities~$\mu$. Therefore, it represents the aggregate traffic
destined to $c$ that could reach node $b(n)$, if }\textcolor{black}{\uline{all}}\textcolor{black}{{}
links were used exclusively and concurrently for this task, using
their full capacity for the time interval $T$. In other words, the
term is generally an upper bound of the traffic incoming to $b(n)$.
Therefore, the focus is on the case where:
\begin{equation}
\underset{l:\,d(l)=b(n)}{\sum}T\left(\mu_{l}^{(c)}+{\color{blue}\mathring{{\color{black}\mu}}_{{\color{black}(n)}}^{{\color{black}(c)}}}\right)>\underset{_{(n,c)}}{U}(t)-\left(\underset{_{(b(n),c)}}{U}(t)+G{}_{(b(n),c)}^{t\to t+T}\right)\label{eq:transitAssumption}
\end{equation}
In this case, the equality in equation (\ref{eq:optimization_goal})
cannot be enforced. However, $RHS_{(\ref{eq:earlyInsights})}$ is
convex, as shown in eq. (\ref{eq:convex}). Therefore, $RHS_{(\ref{eq:earlyInsights})}$
is minimized when its first derivative, i.e., equation (\ref{eq:optimization_goal}),
is closest to zero. This is obtained by maximizing:
\begin{equation}
\Delta_{(n,c)}(t)=\left[U_{(n,c)}(t)-\left(U_{(b(n),c)}(t)+G{}_{(b(n),c)}^{t\to t+T}\right)\right]\label{eq:DeltaU}
\end{equation}
which depends on the traffic generated locally at node $n(b)$ during
$\left[t,t+T\right]$. In other words, the throughput-optimizing routing
decision at node $n$, regarding traffic destined to node $c$ are
derived as follows:
\begin{equation}
n^{*}=argmax_{b(n)}\left\{ \Delta_{(n,c)}(t)\right\} \label{eq:foresightOptimal}
\end{equation}
where $n^{*}$ is the optimal neighboring of $n$ to offload data
to $c$.}

\textcolor{black}{}
\begin{algorithm}[t]
\textcolor{black}{\footnotesize{}\begin{algorithmic}[1]
\Procedure{FBPR}{$network\_state(t=mT|m\in\mathbf{N})$}
\For {each node $n$}
\Comment Define priority flows.
\State $visited[c]\gets 0, \forall c$
\For {each link $l : source(l)=n$}
\State $c_l^{*}(t)\gets \scriptsize{\underset{c:!visited[c]}{argmax}\{U_{(n,c)}-U_{(d(l),c)}-G_{(d(l),c)}^{t\to t+T}\}}$
\State $visited[c_l^{*}(t)]\gets 1$
\State $\Delta Q^{*}_l(t)\gets max\{0,U_{(n,c_l^{*}(t))}-U_{(d(l),c_l^{*}(t))}\}$
\EndFor
\EndFor \Comment Consider multi-links, if any.
\State $\mathbf{\mu}^{*}(t)\gets argmax_{\mathbf{\mu}}\sum_{\forall l}\mu_{l}\cdot \Delta Q^{*}_{l}(t)$
\For {each link $l:\Delta Q^{*}_l(t)>0$}
\State Deploy rule $\{from:s(l),to:c_l^{*}(t),via:l\}$.
\EndFor
\EndProcedure
\end{algorithmic}}{\footnotesize \par}

\textcolor{black}{\footnotesize{}\caption{\label{alg:FBPR}The Foresight-enabled Backpressure Routing algorithm.}
}{\footnotesize \par}
\end{algorithm}

\textcolor{black}{In light of eq. (\ref{eq:foresightOptimal}), we
formulate the Foresight-enabled BPR Routing (FBPR, Algorithm \ref{alg:FBPR}).
The line $5$ of the proposed Algorithm reflects the outcome of equation
(\ref{eq:foresightOptimal}). If an alarm level is defined, the search
in line $5$ is restricted within $c:\,U_{n}^{(c)}\ge alarm\_level$.
The $visited\left[.\right]$ array is also introduced, to make sure
that each possible destination is routed via a single link at each
node. The optimization of line $10$ acquires a different meaning,
pertaining to multi-links. Assume a triple link $M=\left\{ l_{1}:\mu_{1},\,l_{2}:\mu_{2},\,l_{3}:\mu_{3}\right\} $
and a corresponding set of $c_{l}^{*}(t)$ assignments $A=\left\{ c_{l_{1}}^{*}(t),\,c_{l_{2}}^{*}(t),\,c_{l_{3}}^{*}(t)\right\} $.
Line $10$ refers to the optimal reordering of the assignments out
of all possible $M\times A$ combinations and for each multi-link
of the network, maximizing the expected throughput. Finally, lines
11-13 install the FBPR-derived priority rules to the corresponding
nodes.}

\textcolor{black}{Notice that FBPR does not require any critical change
over SBPR. To the contrary, taking into account traffic forecasts
can be achieved by simply introducing the $G$ terms in line~$5$.
Therefore, FBPR does not alter the general workflow of SBPR and retains
the advantage of requiring node-local state information only. In addition,
forecast-awareness can be easily made optional by ignoring $G$ in
line $5$. In this case FBPR falls back to SBPR.}
\begin{cor}
\textcolor{black}{FBPR is throughput-optimal.}
\end{cor}
\textcolor{black}{We notice that the preceding analysis takes place
before the relaxation of equation (\ref{eq:optPursuitBPR}) of the
classic analytical procedure. Applying this final relaxation to equation
(\ref{eq:earlyInsights}) leads to compliance with the Lyapunov stability
criterion (relation (\ref{eq:LyastabilityCriteria})) and to the proof
of throughput optimality, as detailed in \cite{Neely.2005b}. Therefore,
FBPR retains this important trait of SBPR as well.}

\subsection{\textcolor{black}{On the dependence of the proposed system on the
controller.}}

\textcolor{black}{In order to minimize the network's dependency from
the controller and the overhead introduced by the BPR system, the
actuation period $T$ should ideally be as large as possible. However,
FBPR shows that the routing decisions depend on the prediction of
the aggregate traffic that will be generated at each node with the
time interval $[t,t+T]$. Furthermore, the accuracy of predictors
generally decreases as $T$ increases. Thus, it is important to study
the effects of $T$ on the stability of the system.}

\textcolor{black}{Returning to inequality (\ref{eq:RHS_raw}), we
use the theorem of mean value to replace:
\begin{equation}
G_{(n,c)}^{t\to t+T}=\overline{\lambda_{(n,c)}}(t)\cdot T\label{eq:meanValue}
\end{equation}
where $\overline{\lambda_{(n,c)}(t)}$ is the average traffic production
rate at node $n$ towards $c$ for the time interval $[t,t+T]$. Therefore,
we produce:
\begin{multline}
\Delta L(t)\le T^{2}\cdot\underset{_{\forall n,c}}{\sum}\left[{\color{blue}\mathring{{\color{black}\mu}}_{{\color{black}(n)}}^{{\color{black}(c)\,2}}}+\left(\underset{l:\,d(l)=n}{\sum}\mu_{l}^{(c)}+\overline{\underset{_{(n,c)}}{\lambda}}(t)\right)^{2}\right]\\
-2T\cdot\underset{_{\forall n,c}}{\sum}U_{(n,c)}(t)\cdot\left[{\color{blue}\mathring{{\color{black}\mu}}_{{\color{black}(n)}}^{{\color{black}(c)}}}-\underset{l:\,d(l)=n}{\sum}\mu_{l}^{(c)}-\overline{\underset{_{(n,c)}}{\lambda}}(t)\right]\label{eq:RHS_T}
\end{multline}
The RHS of inequality (\ref{eq:RHS_T}) is a quadratic equation of
$T$ of the form $RHS_{(\ref{eq:RHS_T})}(T)=\alpha T^{2}-2\beta T$,
with $\alpha>0$ and roots:
\begin{equation}
\left\{ \begin{array}{c}
\rho_{1}=0\\
\rho_{2}=\frac{2\beta}{\alpha}
\end{array}\right.
\end{equation}
Reminding that $\mu$ are the nominal capacities of }\textcolor{black}{\emph{backbone}}\textcolor{black}{{}
links, it is expected that :
\begin{equation}
\left|\frac{2\beta}{\alpha}\right|\approx0,\,\mbox{for}\,max\left\{ \mu\right\} \ggg1\,\mbox{bps}
\end{equation}
Furthermore, it holds that:
\begin{equation}
\frac{d^{2}RHS_{(\ref{eq:RHS_T})}}{d^{2}T}=2\alpha>0
\end{equation}
Thus, the effect of $T$ on the stability of the network can be summarized
as follows:}
\begin{rem}
\textcolor{black}{\label{lem:The-lower-boundT2}The sensitivity of
the bound of the Lyapunov drift in a BPR-based network is proportional
to the monitoring and actuation period $T$. }
\end{rem}
\textcolor{black}{Lemma \ref{lem:The-lower-boundT2} states that the
increase of $T$ reduces the network stability, for any BPR algorithm
(i.e., both SBPR and FBPR). However, the longevity of routing decisions
may also be equally influenced by the accuracy of the $\overline{\underset{_{(n,c)}}{\lambda}}(t)$
prediction. To demonstrate this fact, we begin by substituting equation~(\ref{eq:meanValue})
in~(\ref{eq:optimization_goal}):
\begin{multline}
T\cdot\left[\underset{l:\,d(l)=b(n)}{\sum}\mu_{l}^{(c)}+{\color{blue}\mathring{{\color{black}\mu}}_{{\color{black}(n)}}^{{\color{black}(c)}}}\right]\\
-\left[U_{(n,c)}(t)-\left(U_{(b(n),c)}(t)+\overline{\underset{_{(b(n),c)}}{\lambda}}(t)\cdot T\right)\right]=0,\,\forall n,c\label{eq:optimization_goal-FOr_L}
\end{multline}
Taking the first derivative of the RHS of equation set (\ref{eq:optimization_goal-FOr_L})
with regard to $\overline{\underset{_{(b(n),c)}}{\lambda}}(t)$ yields:
\begin{equation}
\frac{\partial RHS_{(\ref{eq:optimization_goal-FOr_L})}}{\partial\overline{\underset{_{(b(n),c)}}{\lambda}}(t)}=T
\end{equation}
Reminding that the equation set (\ref{eq:optimization_goal-FOr_L})
yields the optimal routing decisions $b(n)$ we deduce that:}
\begin{rem}
\textcolor{black}{\label{lem:The-sensitivity-of}The sensitivity of
the routing decisions of the FBPR algorithm to the precision of the
$\overline{\underset{_{(n,c)}}{\lambda}}(t)$ predictions is proportional
to $T$.}
\end{rem}
\textcolor{black}{Lemmas \ref{lem:The-lower-boundT2} and \ref{lem:The-sensitivity-of}
both denote the same sensitivity of the system to increases in the
actuation period $T$ and the precision of the $\overline{\underset{_{(b(n),c)}}{\lambda}}(t)$
prediction. Particularly, the introduction of traffic forecasting-awareness
is shown not to introduce a new dominant factor undermining the network
stability. Thus, the choice of $T$ can be made jointly for both factors,
based on the total effect on the Lyapunov drift as follows:}
\begin{defn}
\textcolor{black}{A\label{def:AsoftStaility-1} set of BP routing
decisions is }\textcolor{black}{\uline{acceptable}}\textcolor{black}{{}
for a time period $T_{max}$ such that:
\begin{equation}
RHS_{(\ref{eq:RHS_T})}(T_{max})\le L_{max}
\end{equation}
where $L_{max}$ is a predefined, acceptable constant.}
\end{defn}
\textcolor{black}{Definition \ref{def:AsoftStaility-1} simply ensures
that the Lyapunov drift of the network remains in check for a time
up to $t+T_{max}$, which takes prediction precision errors into account
as well. }

\subsection{\textcolor{black}{On the co-operation of BPR and distance-based routing
schemes.}}

\textcolor{black}{}
\begin{figure}[t]
\begin{centering}
\textcolor{black}{\includegraphics[width=1\columnwidth]{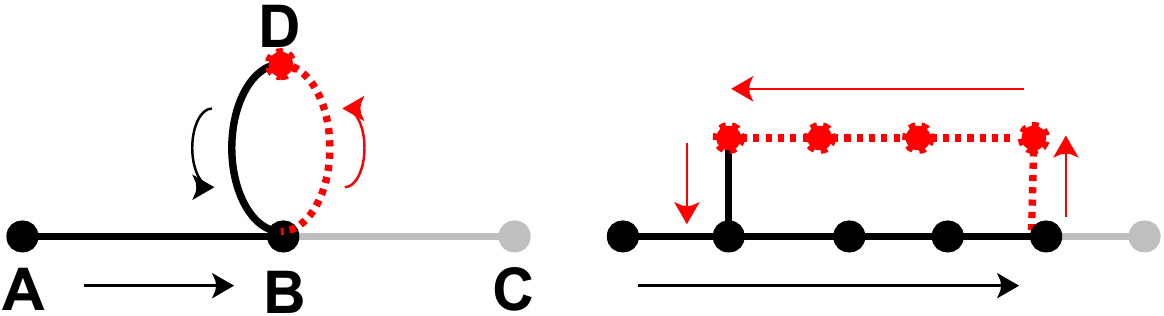}}
\par\end{centering}
\textcolor{black}{\caption{\label{fig:loops}The formation of routing loops is possible when
BPR routing and DVR are naively combined. Solid edges correspond to
DVR rules, while dashed edges designate BPR rules. }
}

\end{figure}

\textcolor{black}{The core assumptions of BPR routing regarding the
queue dynamics, expressed in eq. (\ref{eq:strictQdynamics}), allow
for the incoming, outgoing and generated node traffic to be attributed
to any process. Due to this generality, BPR can operate in parallel
with other data transfer systems, such as DVR, without compromising
its stability and throughput optimality traits. Nonetheless, the loop-free,
latency-aware operation requires further study.}

\textcolor{black}{Pure BPR routing (such as SBPR and FBPR) is orthogonal
to DVR in terms of its optimization goal. BPR is throughput-oriented
and its routing decisions are made on the backpressure potential $\Delta$
of equation~(\ref{eq:DeltaU}). Reducing the number of hops per routing
step is not accounted for. On the other hand, DVR is latency-oriented
and its decisions aim at reducing the network's latency. This conflict
in operational criteria may lead to the formation of routing loops
when DVR and BPR routing decisions are naively combined. Two such
cases are illustrated in Fig.~\ref{fig:loops}. At the left inset,
data flows are generated from node A towards C. A BPR-derived routing
rule intervenes and dictates a detour of the data transfer at node
B towards D. However, the DVR policy of D returns the flows back to
B creating a loop. The same phenomenon can occur after the intervention
of several consecutive BPR rules, as shown at the right inset of Fig.~\ref{fig:loops}. }

\textcolor{black}{We proceed to study the conditions that ensure a
loop-free combination of BPR and DVR policies. The ensuing formulation
will be based on the concept of iterated functions and will refer
to the routing of data towards a given destination node $c$. Let:
\begin{equation}
n'\gets\mathcal{P}(n)
\end{equation}
denote the application of a routing policy $\mathcal{P}$ on data
residing at node $n$, returning their next intended hop $n'$. $\mathcal{P}_{BP}$
denotes the use of BPR, while $\mathcal{P}_{DV}$ the use of DVR.
In addition, let:
\begin{equation}
n'\gets\mathcal{P}^{\circ m}(n)
\end{equation}
denote the $m$-times iterated application of policy $\mathcal{P}$
on node $n$. Finally, $\overrightarrow{T}\left\{ \mathcal{P}^{\circ m}(n)\right\} $
will denote the set of traversed nodes according to policy $\mathcal{P}^{\circ m}$.}
\begin{lem}
\textcolor{black}{A policy $\mathcal{P}$ is loop-free if-and only
if-it has a fixed point on a finite graph.}
\end{lem}
\begin{IEEEproof}
\textcolor{black}{A fixed point of an iterated function $f(.)$ is
a point $x:\,f(x)=x$. Therefore, it holds that $f^{\circ m}(x)=x,\,\forall m$.
Assume that a policy $\mathcal{P}$ has a fixed point and a loop originating
at a node $b$, located $m$ steps away from another node $n$, $b\gets\mathcal{P}^{\circ m}(n)$.
Then, for every $m'>m$, $\mathcal{P}^{\circ m'}(n)$ will cycle over
the nodes comprising the loop, oscillating around $b$, without converging. }
\end{IEEEproof}
\begin{lem}
\textcolor{black}{\label{lem:The LF BP-policy}The $\mathcal{P}_{BP}$
policy is loop-free.}
\end{lem}
\begin{IEEEproof}
\textcolor{black}{Assume an origin node $n_{0}$ and let $\left\{ n_{1},n_{2},\ldots n_{m}\right\} \gets\overrightarrow{T}\left\{ \mathcal{P}^{\circ m}(n_{0})\right\} $.
For each step indexed by $i=0\ldots m$, the BPR algorithm has acted
based on the $\Delta$ metric of equation (\ref{eq:DeltaU}) as follows:
\[
\begin{array}{ll}
i=0: & U_{n_{0}}>U_{n_{1}}+G_{n_{1}}\\
i=1: & U_{n_{1}}>U_{n_{2}}+G_{n_{2}}\\
i=2: & U_{n_{2}}>U_{n_{3}}+G_{n_{3}}\\
\cdots & \cdots
\end{array}
\]
Since the $G$ quantities are strictly positive, it also holds that:
\begin{equation}
U_{n_{i}}+G_{n_{i}}>U_{n_{i+1}}+G_{n_{i+1}},\,i=0\ldots m\label{eq:loco}
\end{equation}
Therefore, the $\overrightarrow{T}\left\{ \mathcal{P}^{\circ m}(n_{0})\right\} $
set is strictly ordered by descending $U_{n_{i}}+G_{n_{i}}$ values.
Assume that $\mathcal{P}_{BP}$ creates a loop originating at some
step $m$, which leads back to a node $n_{j},\,j\le m$ after a number
of steps. This would imply that:
\begin{equation}
U_{n_{m}}+G_{n_{m}}<U_{n_{j}}+G_{n_{j}}
\end{equation}
which contradicts to equation (\ref{eq:loco}). Thus, $\mathcal{P}_{BP}$
must have a fixed point on a finite graph and, therefore, is loop-free.}
\end{IEEEproof}
\textcolor{black}{}
\begin{figure}[t]
\begin{centering}
\textcolor{black}{\includegraphics[width=0.8\columnwidth]{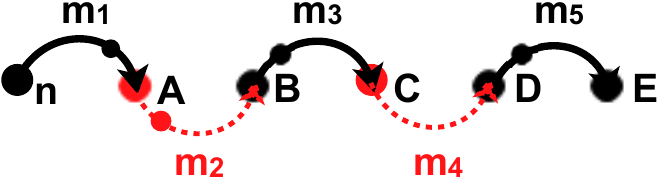}}
\par\end{centering}
\textcolor{black}{\caption{\label{fig:PolicyStitch}A loop-free combination of $\mathcal{P}_{BP}$
and $\mathcal{P}_{DV}$ policies.}
}
\end{figure}

\textcolor{black}{We proceed to study the conditions for loop-free
combinations of $\mathcal{P}_{BP}$ and $\mathcal{P}_{DV}$ policies.
Assume that BPR routing has proposed the path illustrated in Fig.
\ref{fig:PolicyStitch} for routing data flows from an origin node
$n$ towards $E$. Solid and dashed lines correspond to $\mathcal{P}_{DV}$
and $\mathcal{P}_{BP}$ policies respectively, while $m_{x}$ denotes
the number of nodes comprising each separate piece of the path. First,
we note that the path is piece-wise loop-free, since $\mathcal{P}_{BP}$
and $\mathcal{P}_{DV}$ are themselves loop-free. Next, we observe
that the path $n\to B$ is loop-free if:
\begin{equation}
\nexists\mathbf{m}:\,A\in\overrightarrow{T}\left\{ \mathcal{P}_{DV}^{\circ\mathbf{m}}\circ\mathcal{P}_{BP}^{\circ m_{2}}\circ\mathcal{P}_{DV}^{\circ m_{1}}(n)\right\}
\end{equation}
 or equivalently:
\begin{equation}
\nexists\mathbf{m}:\,\mathcal{P}_{DV}^{\circ m_{1}}(n)\in\overrightarrow{T}\left\{ \mathcal{P}_{DV}^{\circ\mathbf{m}}\circ\mathcal{P}_{BP}^{\circ m_{2}}\circ\mathcal{P}_{DV}^{\circ m_{1}}(n)\right\}
\end{equation}
Setting $\mathcal{P}_{CMB}(n)=\mathcal{P}_{DV}^{\circ m_{3}}\circ\mathcal{P}_{BP}^{\circ m_{2}}\circ\mathcal{P}_{DV}^{\circ m_{1}}(n)$
we proceed to check recursively if the following criteria holds:
\begin{equation}
\nexists\mathbf{m}:\,D\in\overrightarrow{T}\left\{ \mathcal{P}_{DV}^{\circ\mathbf{m}}\circ\mathcal{P}_{BP}^{\circ m_{4}}\circ\mathcal{P}_{CMB}(n)\right\} ,\,\text{or}
\end{equation}
\begin{equation}
\nexists\mathbf{m}:\,\mathcal{P}_{BP}^{\circ m_{4}}\circ\mathcal{P}_{CMB}(n)\in\overrightarrow{T}\left\{ \mathcal{P}_{DV}^{\circ\mathbf{m}}\circ\mathcal{P}_{BP}^{\circ m_{4}}\circ\mathcal{P}_{CMB}(n)\right\}
\end{equation}
The process is generalized in the following Lemma:}
\begin{lem}
\textcolor{black}{\label{lem:loopcheck}The introduction of a pathlet
$\mathcal{P}_{BP}^{\circ m}$ at an intermediate node $\mathcal{P}^{*}(n)$
of a loop-free path results into a new loop-free path if it holds
that:
\begin{equation}
\nexists\mathbf{\mathcal{M}}:\,\mathcal{P}_{BP}^{\circ m}\circ\mathcal{P}^{*}(n)\in\overrightarrow{T}\left\{ \mathcal{P}_{DV}^{\circ\mathcal{M}}\circ\mathcal{P}_{BP}^{\circ m}\circ\mathcal{P}^{*}(n)\right\}
\end{equation}
}
\end{lem}
\textcolor{black}{Lemma \ref{lem:loopcheck} provides a quick check
of loop formation when stitching together $\mathcal{P}_{BP}$ and
$\mathcal{P}_{DV}$ policies. Assuming that the topology paths are
cache-able, Lemma \ref{lem:loopcheck} requires a single linear search
per added $\mathcal{P}_{BP}$ pathlet end-point. Furthermore, the
check is additive per every new $\mathcal{P}_{BP}$ pathlet. If the
check yields loop formation, then BPR is run again at the $\mathcal{P}_{BP}$
pathlet start-point, this time excluding the presently selected neighbor
from the search of lines $4-8$, Algorithm \ref{alg:classicBP}. If
no valid neighbor is found, no BPR priority routing rule is installed
at the pathlet start-point. The procedure is formulated as Algorithm
\ref{alg:FBPR-EXHAUSTIVE}. The runtime of the algorithm can be limited
by an allowed $Timeout$ parameter, expressing a potential deadline
imposed by the network regarding the derivation of the priority rules.}
\begin{algorithm}[!t]
\textcolor{black}{\footnotesize{}\begin{algorithmic}[1]
\Procedure{BP\_DV\_Stitch}{$state_{netw}(t),Timeout$}
\For {\textbf{each} node $n$}
\State $neighbors[n]\gets \{\text{nodes }n': \exists\text{ link }l_{n\to n'}\}$.
\EndFor
\State $Loop\_found\gets true$.
\State \textbf{while} $Loop\_found$ \textbf{and} $Timeout$ not exceeded,
\State\hspace{\algorithmicindent} $Loop\_found\gets false$.
\State\hspace{\algorithmicindent} Execute $FBPR(state_{netw}(t), neighbors)$.
\State\hspace{\algorithmicindent} Detect loops via Lemma \ref{lem:loopcheck}.
\State\hspace{\algorithmicindent} \textbf{for each} loop-inducing BP pathlet $p$ \textbf{do}
\State\hspace{\algorithmicindent}\hspace{\algorithmicindent} $Loop\_found\gets true$.
\State\hspace{\algorithmicindent}\hspace{\algorithmicindent} $neighbors[startpoint\{p\}]-=current\_neighbor$.
\State\hspace{\algorithmicindent} \textbf{end for}
\hspace{\algorithmicindent}\State \textbf{end while}
\State Deploy loop-free BP priority rules.
\EndProcedure
\end{algorithmic}}{\footnotesize \par}

\textcolor{black}{\footnotesize{}\caption{\label{alg:FBPR-EXHAUSTIVE}The FBPR and DVR policy stitching algorithm.}
}{\footnotesize \par}
\end{algorithm}

\textcolor{black}{The $\mbox{BP\_DV\_STITCH}$ algorithm is intended
to promote the exploration of all possible neighbor choices that may
offer an alternative exit to locally accumulated traffic via BPR.
This exploratory nature enforces the loop control of line~$9$, as
well as the iteration of lines~$6-14$. The exploration depth of
the algorithm can also be limited deterministically. This choice can
offer immunity against issues of asynchronous installation of flow
rules, which presently constitutes a significant limitation of OpenFlow-based
solutions \cite{Reitblatt.2011b,Reitblatt.2012,McGeer.2012,Katta.2013}.
This approach is expressed via Algorithm \ref{alg:FBPR-NHOPS}.}
\begin{algorithm}[!t]
\textcolor{black}{\footnotesize{}\begin{algorithmic}[1]
\Procedure{NHOPS\_Stitch}{$state_{netw}(t)$}
\For {\textbf{each} origin node $n$}
\For {\textbf{each} destination node $c\ne n$}
\State $n'\gets \{n: \left\Vert \overrightarrow{T}_{(c)}\left\{ \mathcal{P}_{DV}^{\circ\infty}(n)\right\} \right\Vert >\left\Vert \overrightarrow{T}_{(c)}\left\{ \mathcal{P}_{DV}^{\circ\infty}(n')\right\} \right\Vert\}$.
\State $neighbors[n][c]\gets n'$.
\EndFor
\EndFor
\State Execute $FBPR(state_{netw}(t), neighbors)$.
\EndProcedure
\end{algorithmic}}{\footnotesize \par}

\textcolor{black}{\footnotesize{}\caption{\label{alg:FBPR-NHOPS}Less exploratory FBPR and DVR policy stitching.}
}{\footnotesize \par}
\end{algorithm}

\textcolor{black}{Inspired by the study of Ying et al.~\cite{Ying.2011b},
Algorithm~\ref{alg:FBPR-NHOPS}, at line 4, limits the BPR search
space of neighbors only to those that offer a decreased number of
hops towards the destination $c$. This approach disallows the formation
of routing loops, as expressed by the following Lemma: }
\begin{lem}
\textcolor{black}{\label{lem:-is-loop-free.}$\mbox{NHOPS\_Stitch}$
is loop-free.}
\end{lem}
\begin{IEEEproof}
\textcolor{black}{Consider any path of arbitrary length $m$ from
a node $n$ to $c\ne n$, $\left\{ n_{1},n_{2},\ldots,n_{m}\right\} =\overrightarrow{T}_{(c)}\left\{ \mathcal{P}^{\circ m}(n)\right\} $,
where $\mathcal{P}^{\circ m}$ comprises any combination of $\mathcal{P}_{BP}$
and $\mathcal{P}_{DV}$ policies. Due to line 4 of Algorithm \ref{alg:FBPR-NHOPS},
the nodes $\left\{ n_{1},n_{2},\ldots,n_{m}\right\} $ are order by
descending number of hops towards the destination $c$. As in Lemma
\ref{lem:The LF BP-policy}, the existence of a loop would violate
this ordering, concluding the proof. }
\end{IEEEproof}
\textcolor{black}{Finally, it holds that:}
\begin{lem}
\textcolor{black}{$\mbox{NHOPS\_Stitch}$ requires no mechanism to
guard against asynchronous or partial deployment of the BP priority
rules in the network.}
\end{lem}
\begin{IEEEproof}
\textcolor{black}{The claim is proved by induction. Assume a loop-free
path of arbitrary length $m$ from a node $n$ to $c\ne n$, $\left\{ n_{1},n_{2},\ldots,n_{m}\right\} =\overrightarrow{T}_{(c)}\left\{ \mathcal{P}^{\circ m}(n)\right\} $,
where $\mathcal{P}^{\circ m}$ comprises any combination of $\mathcal{P}_{BP}$
and $\mathcal{P}_{DV}$ policies. We install an additional priority
flow rule on any of the nodes $n_{i}$ that are governed by $\mathcal{P}_{DV}$
flow entries only. The resulting path $\overrightarrow{T}_{(c)}'\left\{ \mathcal{P}^{\circ m}(n)\right\} $
is still loop-free due to Lemma \ref{lem:-is-loop-free.}. However,
there certainly exists at least one loop-free path, namely the $\overrightarrow{T}_{(c)}\left\{ \mathcal{P}_{DV}^{\circ m}(n)\right\} $,
thus proving the claim. }
\end{IEEEproof}

\section{\textcolor{black}{Simulations\label{sec:Evaluation}}}

\textcolor{black}{In this Section, the performance of the proposed
scheme is evaluated in terms of achieved average throughput, latency
and traffic overflow rate in a variety of settings. The employed simulator
(implemented on the AnyLogic platform (JAVA) \cite{XJTechnologies.2013})
and datasets are freely available.}

\textcolor{black}{The simulations evaluate the potential of collaboration
among a cluster of existing ASes. To this end, a real topology of
ASes covering the European continent is derived, using the Macroscopic
Internet Topology Data Kit (ITDK) by the Center for Applied Internet
Data Analysis (CAIDA)~\cite{CAIDAIDTK.2016}. ITDK offers a router-level
topology of the Internet. Each router is accompanied by its connectivity
(links), its geo-location (city-level) and its AS assignment. From
this complete dataset, we derive a simulation-wise tractable subset
of $25$ ASes. This upper bound was selected via runtime trials in
the simulation platform. The studied subset is derived as follows.
}
\begin{figure}[t]
\begin{centering}
\textcolor{black}{\includegraphics[bb=0bp 0bp 708bp 650bp,width=0.7\columnwidth]{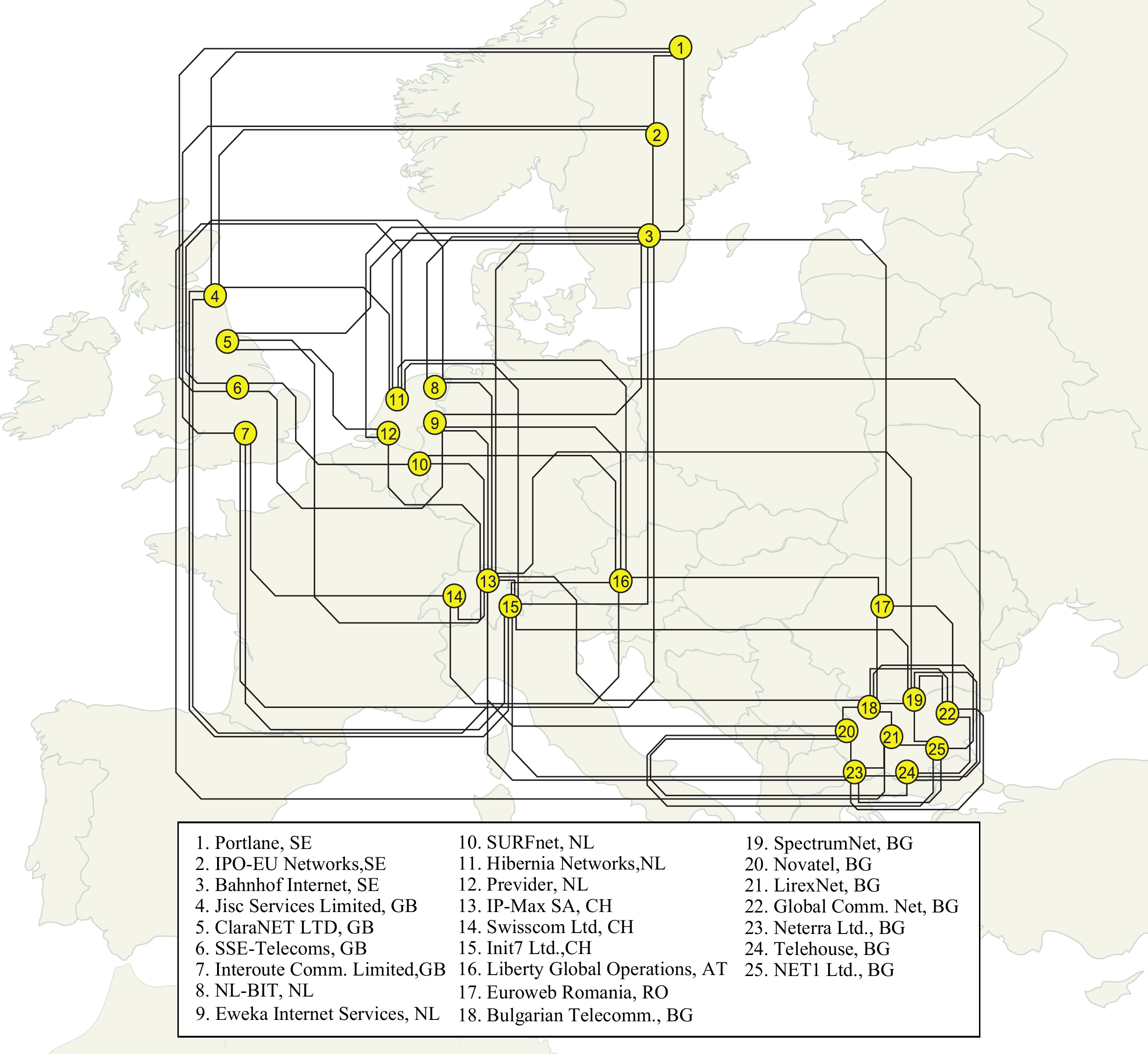}}
\par\end{centering}
\textcolor{black}{\caption{\label{fig:topoAS}\textcolor{black}{The AS-level topology employed
in the simulations, comprising 25 ASes and 66 peering relations. Each
node represents an AS and is annotated with its name and location
(country code within Europe).}}
}
\end{figure}
\textcolor{black}{}
\begin{figure}[t]
\begin{centering}
\textcolor{black}{\includegraphics[bb=0bp -1bp 517bp 451bp,width=0.73\columnwidth]{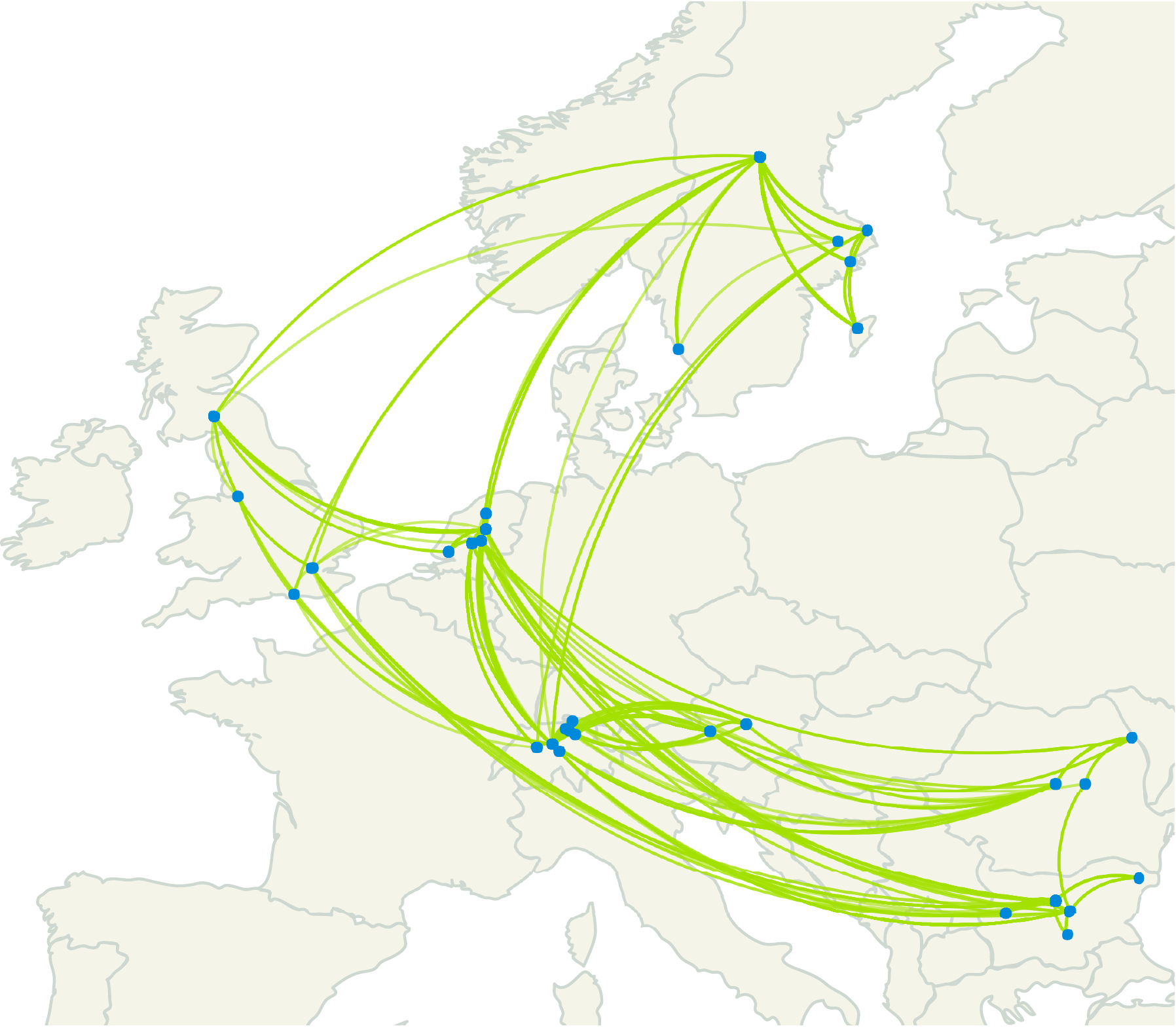}}
\par\end{centering}
\textcolor{black}{\caption{\label{fig:topoRoute}\textcolor{black}{The router-level topology
corresponding to Fig~\ref{fig:topoAS}, containing 351 routers and
273 peering links. (Several router geo-locations coincide at city-level
granularity).}}
}
\end{figure}

\textcolor{black}{Initially, all routers in the ITDK set are mapped
to their corresponding ASes. Subsequently, routers that belong to
non-Transit ASes are filtered out, using the AS classification dataset
by CAIDA~\cite{CAIDAASClassification.2016}. Routers assigned to
Tier-1 ASes are discarded as well, since collaboration between such
organizations is not straightforward~\cite{Renesys.2015}. The resulting
router/AS entries are filtered based on their ITDK geolocation, focusing
on Europe. Moreover, we retain routers/links that are parts of AS
peer-to-peer relations only, using the AS relationships dataset by
CAIDA~\cite{CAIDAASREL.2016}. We sort the remaining ASes by connectivity
degree and keep the $25$ top-connected ones. The final AS-level topology
is illustrated in Fig.~\ref{fig:topoAS}, accompanied by the AS company
names, derived via a CIDR report~\cite{CIDRASLOCATIONS.2016}. The
corresponding router-level topology is illustrated in Fig.~\ref{fig:topoRoute}. }

\textcolor{black}{It is noted that the employed datasets may contain
minor inconsistencies, as reported by the respective publishers. However,
the quality of the employed datasets is considered adequate for the
scope of the evaluation, which focuses on the efficiency of the proposed
inter-AS routing scheme and not on contributing Internet measurements.
Additionally, assumptions are needed in the place of certain attributes
that are not publicly available. First, the physical or virtual (e.g.,
remote peering) nature of the links is omitted in the used datasets.
Nonetheless, the proposed scheme can operate on any link, regardless
of its nature, provided that an AS can freely deploy priority routing
rules that refer to it. Second, the ITDK router topology refers to
AS border routers, but not to their internal connectivity. Therefore,
using a common assumption~\cite{FULLMESHIBGP}, we consider full-mesh
internal AS connectivity. Third, the latency and capacity of Internet
links are generally not publicly available. Thus, we assume that the
latency of a link is derived by the distance between its endpoints,
divided by the speed of light ($3\cdot10^{8}$ m/s). Finally, noticing
that $10$ Gbps rates are commonly supported by CISCO border routers~\cite{CISCOASR.2015},
we pick the capacity of each link at random (uniformly) within the
range $\left[5,\,15\right]$ Gbps at each direction. Multiple runs
with different link capacities are executed.}

\textcolor{black}{The Controller is placed at the Swiss AS ``\#13''
in Fig.~\ref{fig:topoAS}, which yields the highest connectivity
degree in the examined AS graph. AS routers nearest to the Controller
in terms of hops act as NCS endpoints. At the simulation initialization
stage, an instance of the Bellman-Ford algorithm runs at each router,
deriving the underlying DVR routing rules. Subsequently, the Controller
begins to interact with the NCSes with a period of $T$ sec. All installed
priority rules are retained for a full period. Notice that control
messages (i.e., affecting the communication of the Controller and
the NCSes) receive top routing priority.}

\textcolor{black}{Regarding the internal architecture of a router,
the end-points of each router link are connected to router network
interfaces (NICs). The aggregate incoming traffic at each router (from
all NICs) is first enqueued at a central, shared memory ($4$ GB~\cite{CISCOASR.2015}),
and is subsequently dispatched to the appropriate exit NIC based on
the active routing rules. Each NIC is equipped with a $50$ MB-sized
twin-buffer to avoid idle intervals. }

\textcolor{black}{The connection of each border router to the cluster-external
Internet is represented by a dedicated network interface. Each router
hosts one such NIC, which is connected to an instance of an inter-domain
traffic generator (ITMGen)~\cite{ITMGEN}. All such instances within
the same AS are identical, and their cumulatively produced traffic
rate complies with ITMGen, which specifies a traffic matrix describing
the average traffic flow between any AS pair. ITMGen requires as input
a metric of mutual popularity, $p_{ij}$, between any two ASes $i,\,j$,
which describes the portion of traffic that enters~$i$ (cluster-external),
destined towards~$j$. In absence of publicly available data, we
derive $p_{ij}$ from the connectivity degree $d_{i}$ of the ASes
in the setup of~Fig.~\ref{fig:topoAS}. In other words, we assume
that the connectivity degree of an AS reflects its popularity in terms
of serving as traffic endpoint. Initially, the aggregate popularity
$p_{i}$ of an AS $i$ is derived as $p_{i}=\frac{d_{i}}{\sum_{\forall i}d_{i}}$.
Subsequently, $p_{ij}$ is approximated as $p_{ij}=\frac{p_{j}}{1-p_{i}},\,i\ne j$
and $p_{ii}=0$. All other configuration parameters of ITMGen are
retained~\cite{ITMGEN}. Given that packet-level simulation of backbone
networks is generally not tractable in terms of simulation runtimes~\cite{AlFares.2010,Curtis.2011b},
we assume that the generated traffic is organized in $50$ MB-sized
batches (i.e., equal to the NIC twin-buffer size).}

\textcolor{black}{The logged metrics include the network-wide throughput,
overflow rate and average latency. Throughput is calculated as the
total traffic volume that has traversed the network links during the
simulation, divided by the simulation duration. The overflow metric
is defined in a similar fashion, while the latency is measured as
the average delivery time of packet-batches. Finally, in the ensuing
Figures each separate plot is distinguished by the employed BRP variant
(FBPR or SBPR), the DVR-BPR co-existence approach (STITCH for Algorithm~\ref{alg:FBPR-EXHAUSTIVE}
and NHOPS for Algorithm~\ref{alg:FBPR-NHOPS}). FBPR incorporates
a simple averaging forecasting method with window size equal to $T$.
Each simulation lasts for $1$ hour, which was observed to be sufficient
for more than $95\%$ confidence in the logged metrics.}

\subsection{\textcolor{black}{Results}}

\begin{figure}[t]
\begin{centering}
\textcolor{black}{}\subfloat[\label{fig:THPUTpercentT10}\textcolor{black}{Throughput performance.}]{\begin{centering}
\textcolor{black}{\includegraphics[bb=0bp 0bp 420bp 173bp,clip,width=1\columnwidth]{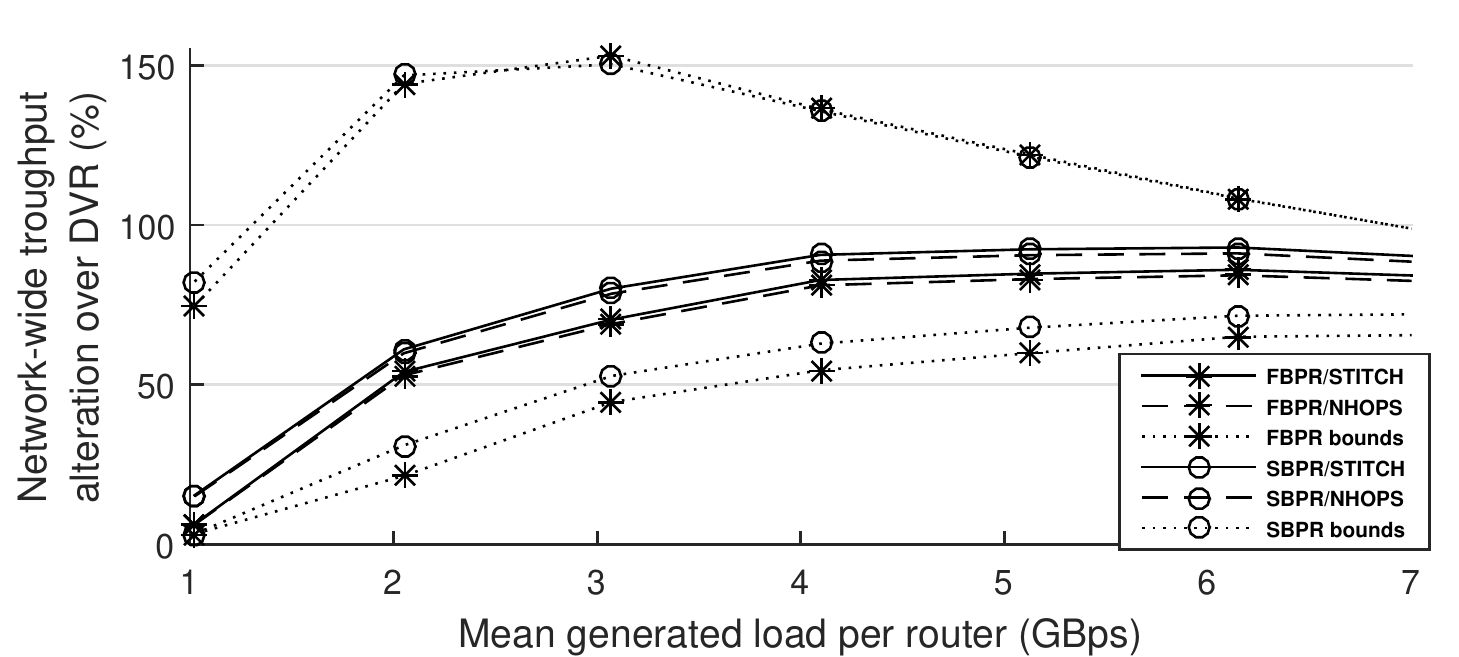}}
\par\end{centering}
\textcolor{black}{}}\textcolor{black}{\vspace{-11bp}
}\subfloat[\label{fig:OVFpercentT10}\textcolor{black}{Overflow performance.}]{\begin{centering}
\textcolor{black}{\includegraphics[bb=0bp 0bp 420bp 188bp,width=1\columnwidth]{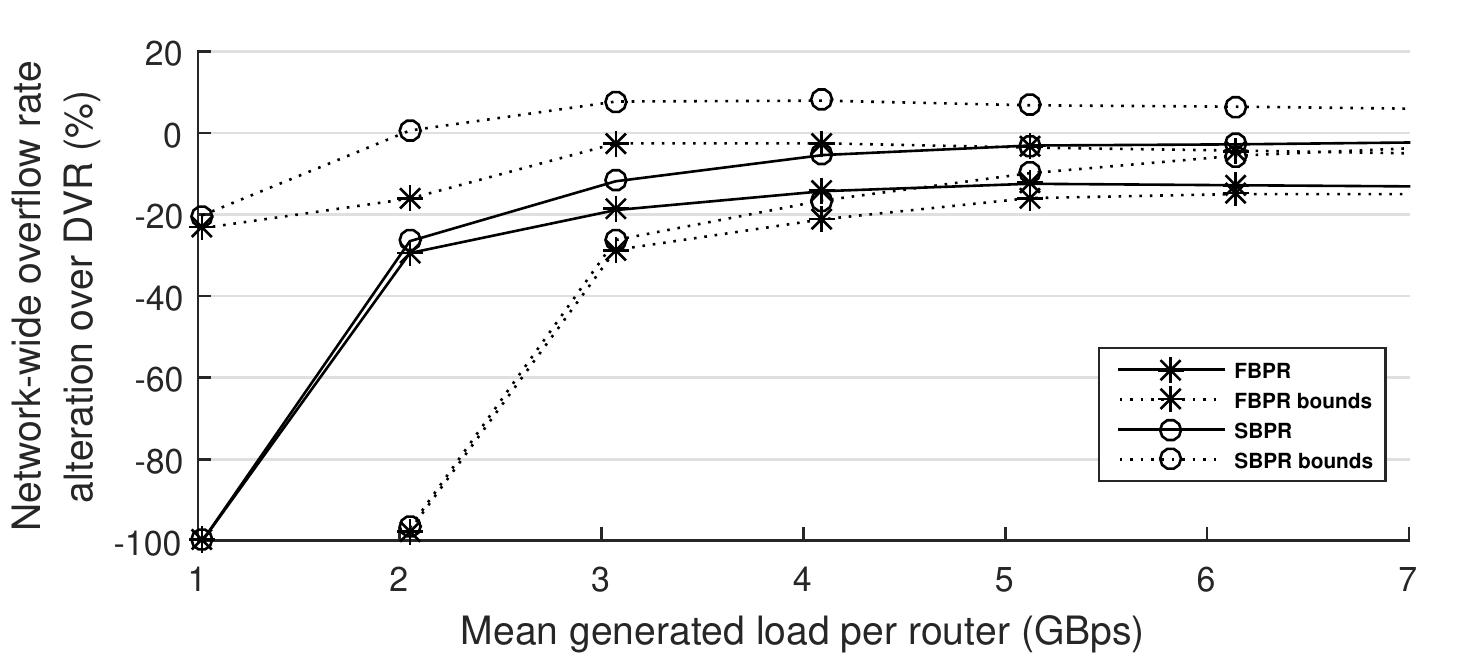}}
\par\end{centering}
\textcolor{black}{}}\textcolor{black}{\vspace{-11bp}
}\subfloat[\label{fig:LATpercentT10}\textcolor{black}{Batch delivery times.}]{\begin{centering}
\textcolor{black}{\includegraphics[width=1\columnwidth]{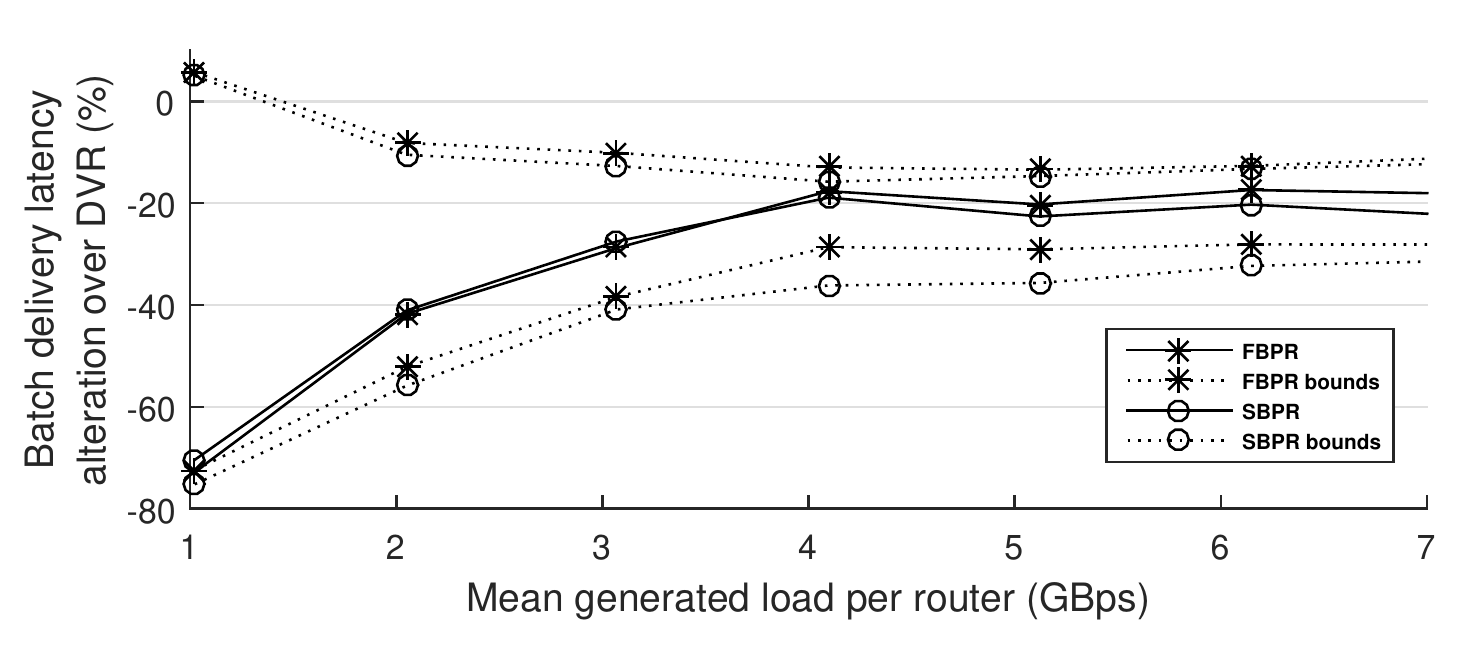}}
\par\end{centering}
\textcolor{black}{}}
\par\end{centering}
\textcolor{black}{\caption{\label{fig:EvalTOLvsLinG}\textcolor{black}{Evaluation of the proposed
scheme versus plain DVR ($T=10$sec).}}
}
\centering{}\textcolor{black}{\vspace{-16bp}
}
\end{figure}

\textcolor{black}{Figure~\ref{fig:EvalTOLvsLinG} evaluates the traffic
volume increase potential of the proposed scheme. To this end, the
actuation period, $T$, is set to a value of $10$ sec, given that
the AS-state monitoring itself requires $3-5$ sec in a real system~\cite{Tootoonchian.2010}.
The ITMGEN traffic matrix is scaled linearly, creating the load axis
values. Boundary and average values over $100$ link capacity randomizations
are shown.}
\textcolor{black}{{} }

\textcolor{black}{According to Fig.~\ref{fig:THPUTpercentT10}, the
proposed scheme can increase the average traffic volume by $100\%$
over the plain DVR case. The lower bound is close to the average values,
while the upper bound reaches up to $150\%$. The STITCH-based variations
combinations achieve slightly better performance than the NHOPS-based
ones. This outcome is expected, given that the NHOPS approach restricts
the AS neighborhood search within decreasing hop distances to the
end destinations. Naturally, the possible choices for traffic offloading
are reduced, which affects the total throughput. Nonetheless, NHOPS
retains a high throughput performance, coupled with natural loop-free
operation benefits.}

\textcolor{black}{The stability of the collaboration cluster also
benefits from the proposed scheme, as shown in Fig.~\ref{fig:OVFpercentT10}.
All proposed variations achieve lower average overflow rates than
plain DVR. The FBPR variations achieve better results than SBPR, with
regard to both average and boundary values. Notice that SBPR yielded
marginally better throughput performance in Fig.~\ref{fig:THPUTpercentT10}.
Nonetheless, this surplus is overflown, since it can be directed to
rapidly overloading ASes. FBPR achieves a better management of traffic
by considering future traffic levels. }

\textcolor{black}{The batch latency also benefits from the proposed
scheme, as shown in Fig.~\ref{fig:LATpercentT10}. This observation
holds for all examined scheme variations. BPR is known to minimize
the amount of queued traffic by distributing it fairly across all
collaborating nodes. As a result, the queuing time decreases as well. }

\textcolor{black}{The ensuing experiments assume the NHOPS variation
only (STITCH is similar) and a medium router load of $4$ GBps. }
\begin{figure}[t]
\begin{centering}
\textcolor{black}{\includegraphics[width=0.33\columnwidth]{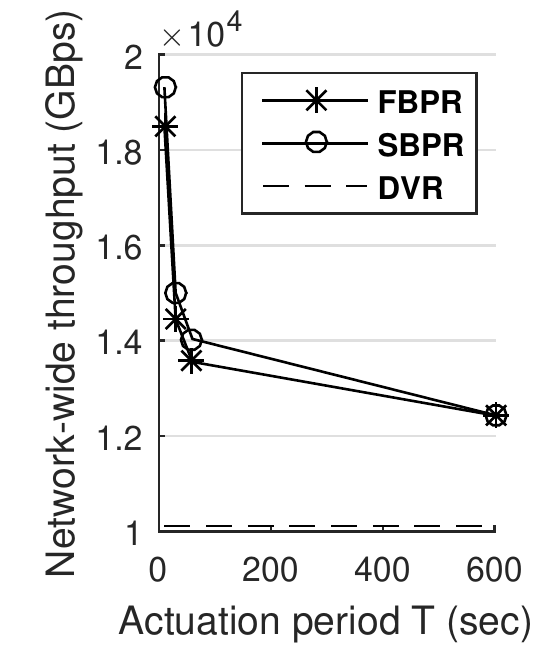}\includegraphics[width=0.33\columnwidth]{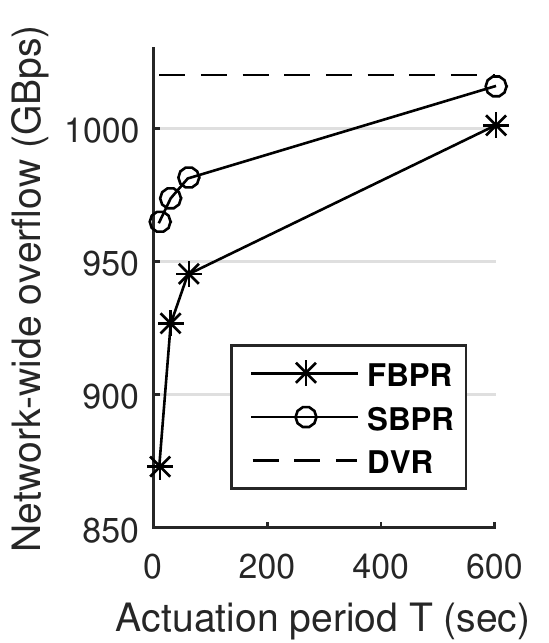}\includegraphics[width=0.33\columnwidth]{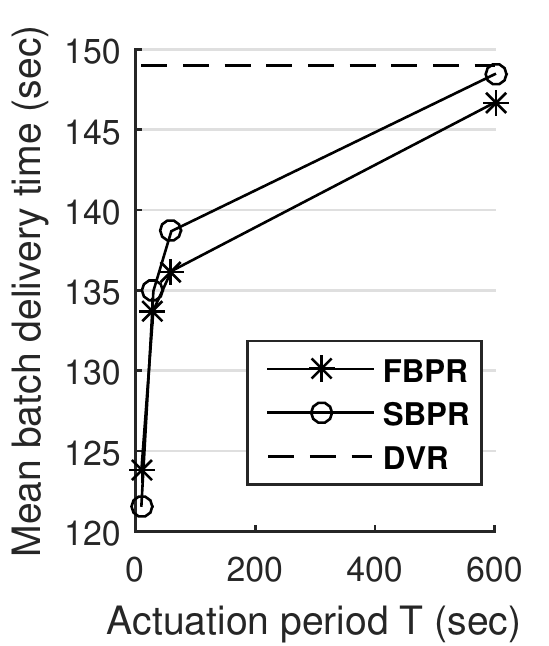}}
\par\end{centering}
\textcolor{black}{\caption{\label{fig:T_Effect}\textcolor{black}{Performance effect of the period
$T$. }}
\vspace{-12bp}
}
\end{figure}

\textcolor{black}{We proceed to study the effects of the actuation
period on the performance of the proposed scheme. Figure~\ref{fig:T_Effect}
presents absolute performance results for periods ranging from $10$
sec to $10$ min. The achieved throughput (left inset) is consistently
better than plain DVR, even for the maximal $T$ value. The throughput
naturally decreases as $T$ increases, given that the derived priority
rules will generally lose their timeliness. The overflow and latency
performance is given in the middle and right insets respectively.
The relative ranking of the compared schemes is retained. Increasing
the actuation period limits the stability and latency benefits of
the proposed scheme as well, retaining an improvement over DVR nonetheless. }

\textcolor{black}{Given that FBPR and SBPR are both throughput-optimal,
their close performance in Fig.~\ref{fig:T_Effect} (left) is expected.
However, in terms of stability (middle), the gains of FBPR are clear
(followed by its latency performance - right). Notice that FBPR at,
e.g., $T=600$ sec behaves as SBPR at $T=350$ sec. Thus, ASes can
also use forecasting for increasing their Controller interaction period,
while retaining SBPR-level stability. }
\begin{figure}[t]
\begin{centering}
\textcolor{black}{\includegraphics[width=1\columnwidth]{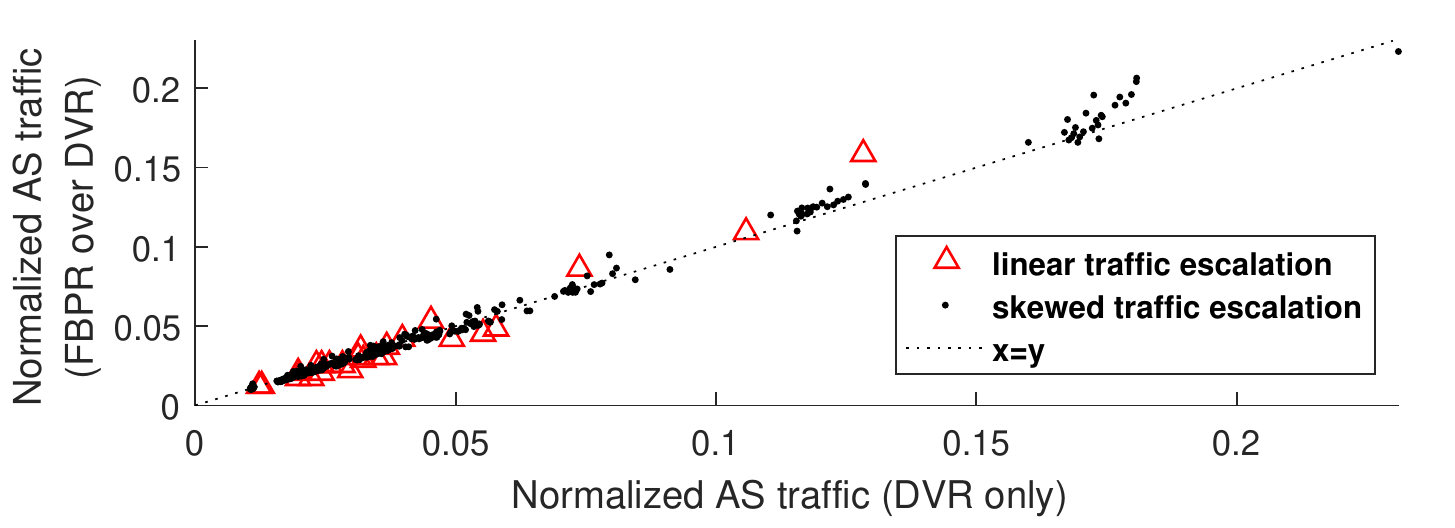}\vspace{-23bp}
}
\par\end{centering}
\textcolor{black}{\caption{\label{fig:DISTRO}\textcolor{black}{Distribution of surplus traffic
to each AS, with and without FBPR.}}
\vspace{-10bp}
}
\end{figure}

\textcolor{black}{We proceed to study the distribution of the transit
traffic over the collaborating ASes. We perform runs with different
cases of router load escalation and log the throughput traffic percentage
per AS under: i) plain DVR and ii) FBPR/NHOPS/T=10sec over DVR. The
}\textcolor{black}{\emph{linear}}\textcolor{black}{{} escalation case
is the proportional escalation of the ITMGEN traffic matrix leading
to $4$ GBps average router load. In the }\textcolor{black}{\emph{skewed}}\textcolor{black}{{}
escalation case, the same traffic load enters the network, but solely
from one given AS. In the skewed case, $25$ separate runs are executed,
with one AS acting as the traffic input point. The ensuing throughput
traffic distribution assumes the non-input ASes. The DVR and FBPR/DVR
results are given as a scatter plot in Fig.~\ref{fig:DISTRO}. All
skewed-case runs are collectively plotted.}

\textcolor{black}{Figure~\ref{fig:DISTRO} exhibits a strongly linear
relation between the traffic distribution over plain DVR, and the
corresponding distribution when FBPR is activated. Thus, the traffic
increase due to FBPR at each AS is fair and proportional to its original
DVR traffic, in accordance with BPR theory (cf. end of Section~\ref{sec:Prerequisites}).
In other words, FBPR will generally not alter the original AS ranking
in terms of served traffic volume. The SBPR behavior is similar and
is omitted for Figure clarity. }
\begin{figure}[t]
\begin{centering}
\textcolor{black}{\includegraphics[width=0.45\columnwidth]{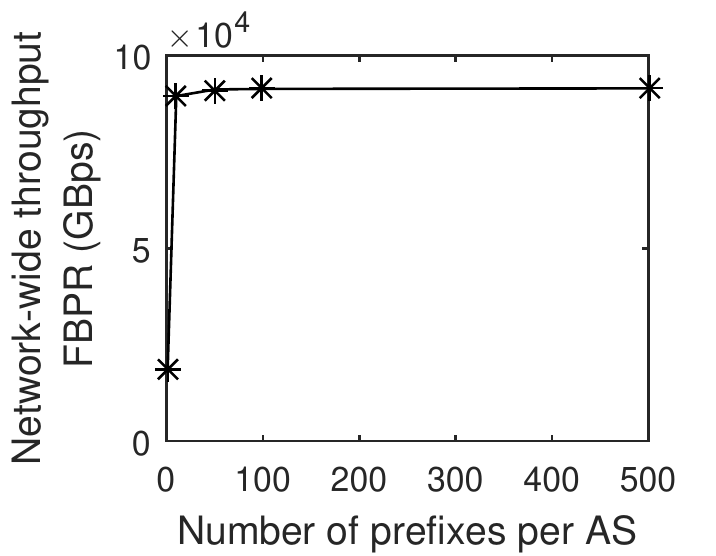}\includegraphics[width=0.45\columnwidth]{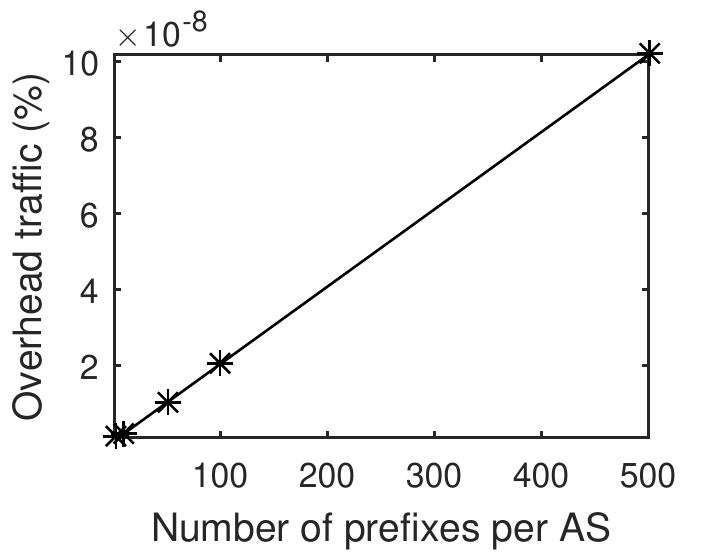}}
\par\end{centering}
\textcolor{black}{\caption{\label{fig:PREFIX}\textcolor{black}{FBPR operation at prefix-level
and incurred overhead.}\textcolor{blue}{{} }}
\vspace{-16bp}
}
\end{figure}

\textcolor{black}{So far, the evaluation assumed traffic routing at
AS-granularity. Figure~\ref{fig:PREFIX} studies the effect of routing
at prefix-granularity on the proposed scheme. We focus on FBPR (SBPR
is similar), using the default mean router load ($4$ GBps) and $T=10$
sec. Each AS hosts a varying number of prefixes $N_{P}$ (x-axis of
Fig.~\ref{fig:PREFIX}), and uses separate backlogs ($U(t)$) for
each one. The FBPR operation follows the steps of Algorithm~\ref{alg:FBPR},
with the modification that $\nicefrac{N_{P}}{N_{L}}$ priority rules
can be assigned to each router link, $N_{L}$ being the number of
outgoing router links. Moreover, the network overhead imposed by FBPR
is logged, assuming the worst case scenario where each AS reports
its full internal congestion per prefix every $T=10$ sec. The report
is a set comprising the AS identifier (once, 8-byte string), prefix
IPs ($4$ bytes each), prefix subnets ($1$ byte each) and prefix
load in GB ($8$-byte float each). A priority rule proposal is a tuple
comprising the intended AS identifier (once), and the prefix IPs ($4$
bytes), subnets ($1$ bytes) and link identifiers ($8$-byte string).}

\textcolor{black}{As shown in Fig.~\ref{fig:PREFIX} (left inset),
per-prefix routing naturally increases the achieved network throughput,
since it allows for more fine-grained traffic management compared
to per-AS routing. Even for a small number of prefixes ($N_{p}=10$),
the performance of FBPR increases by a factor of $\times5$, essentially
reaching the maximal value. The imposed overhead is trivial (Fig.~\ref{fig:PREFIX}
- right inset), amounting at $10^{-8}\%$ of the overall traffic volume,
while scaling linearly with the number of prefixes.}

\textcolor{black}{We note that, in Fig.~\ref{fig:PREFIX}, all ASes
are assumed to manage network prefixes in the same manner. In reality,
an AS$_{1}$ may treat several prefixes individually, while a neighbor
AS$_{2}$ may treat them as a single super-prefix. Both SBPR and FBPR
can operate without changes in the AS prefix management. When considering
backlog differences, SBPR and FBPR simply then consider the affected
backlogs at each side. E.g., offloading AS$_{2}$ to AS$_{1}$ will
consider the backlog difference of the load for the super-prefix at
AS$_{2}$ minus the cumulative load of all affected sub-prefixes at
AS$_{1}$.}

\textcolor{black}{}
\begin{figure}[t]
\begin{centering}
\textcolor{black}{\includegraphics[width=0.33\columnwidth]{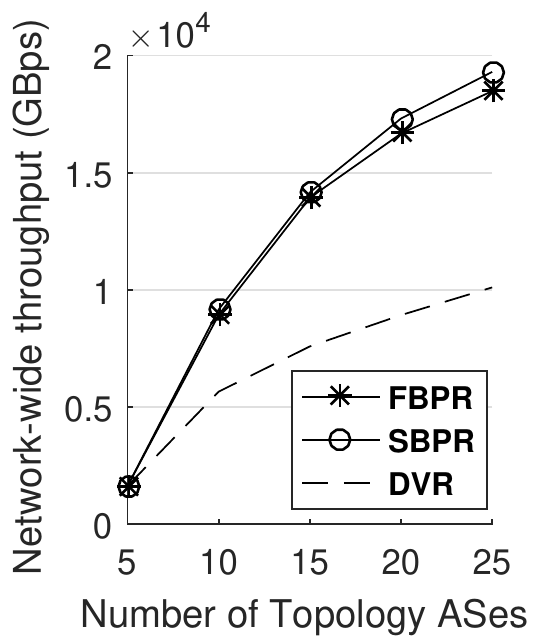}\includegraphics[width=0.33\columnwidth]{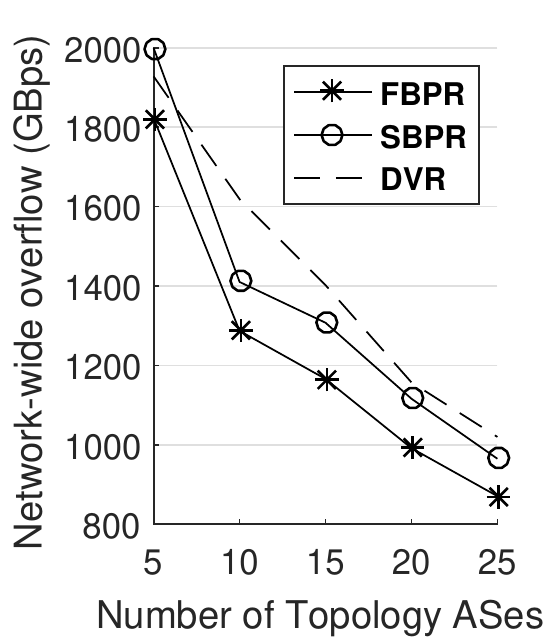}\includegraphics[width=0.33\columnwidth]{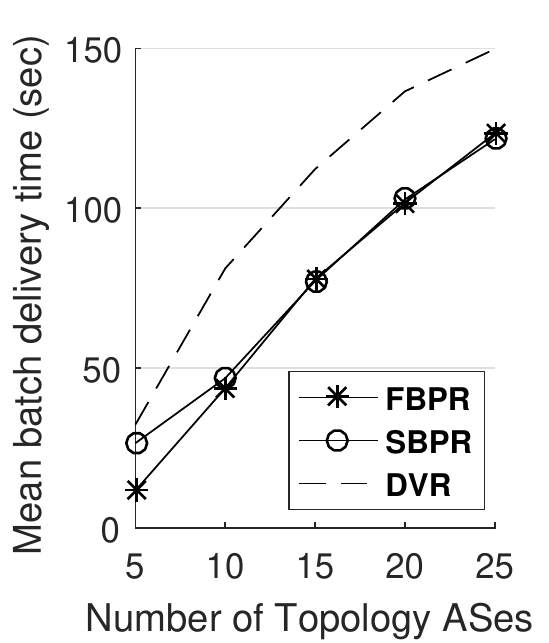}}
\par\end{centering}
\textcolor{black}{\caption{\textcolor{blue}{\label{fig:N_Effect}}\textcolor{black}{Performance
effect of the topology size.}}
\vspace{-16bp}
}
\end{figure}

\textcolor{black}{Finally, Fig.~\ref{fig:N_Effect} studies the effect
of topology size on the performance of the proposed scheme. We assume
the default router load ($4$ GBps), $T=$$10$ sec and per-AS routing.
We select size-varying subsets of the original 25 ASes (producing
the corresponding router-level topologies as well), forming the x-axis
of Fig.~\ref{fig:N_Effect}. Link capacities are randomized $100$
times, presenting average performance values in the y-axis.}

\textcolor{black}{Naturally, vary small topologies (5 ASes) offer
limited path diversity and network capacity. The paths are shorter
nonetheless, yielding decreased latency. However, at slightly bigger
topologies (10 ASes), the throughput, stability and latency benefits
of backpressure appear. The improvement over plain DVR then increases
with the topology size, as new paths are added and exploited by the
BPR schemes. \vspace{-8bp}
}

\subsection{\textcolor{black}{Discussion and Future Steps}}

\textcolor{black}{The evaluation focused on factors that can constitute
the proposed scheme appealing for initiating inter-AS collaboration.
Specifically, three benefits were highlighted: i) Attaining a significant
increase in the transit traffic volume handled by the AS-cluster.
ii) A natural gain-sharing mechanism among the collaborating ASes.
Each particular AS sees an increase in its own transit traffic based
on its centrality within the AS-topology. iii) Lightweight implementation,
requiring a web service accessible by the ASes. The point-of-failure
risk is expected to be low, given that the scheme then naturally falls
back to standard AS routing. Emphasis is placed on using the NCSes
that ASes have already adopted. This means that the proposed routing
rules are not deployed automatically by the proposed scheme, but rather
passed to each NCS for further processing without overriding it. This
approach allows the ASes to retain the full control of their networks. }

\textcolor{black}{Nonetheless, the proposed scheme assumes ASes willing
to cooperate. Trust and information exchange are sensitive matters
in commercial relations. The proposed scheme requires ASes to trust
it with partial information on their internal congestion state. For
instance, an AS$_{1}$ may inform the proposed scheme that ``$10$
GB are pending towards prefix $p$''. This information exchange requires
a degree of trust. Nonetheless, ASes can already obtain similar information
via measurements. For example, an AS$_{2}$ that neighbors AS$_{1}$
can easily derive that ``$10$ GB were received from AS$_{1}$ towards
prefix $p$'' and yield a forecast on the present state of AS$_{1}$.
Thus, the present scheme mainly requires a change in the way such
information is exchanged and not in its confidentiality. }

\textcolor{black}{Future work is directed towards the important aspect
of security against AS foul play, and particularly against reporting
false congestion levels $U_{(n,c)}(t)$. The concept of promise/deliverable
can constitute the basis for detecting such attempts. For example,
if an AS reports falsely high $U_{(n,c)}(t)$ levels to elicit assistance,
it should also deliver this amount of real traffic. This can be verified
via measurements by the assisting ASes. In cases where falsely low
$U_{(n,c)}(t)$ levels are reported to attract traffic, the attracting
AS should yield low latency and loss rate in handling this traffic.
Failure to deliver can imply foul-play, evicting untrustworthy ASes
from the cluster.\vspace{-8bp}
}

\section{\textcolor{black}{Related Work\label{sec:Related-Work}}}

\textcolor{black}{Highly-efficient and even optimal TE within datacenters
constituted one of the earliest successes of SDN. Emphasis was placed
on versatile traffic prioritization systems at flow granularity, as
well as network-wide optimization objectives such as throughput maximization.
B4 incorporates this concern by keeping tuples of source, destination
and QoS traits per network flow~\cite{Jain.2013}. The network's
resources are constantly monitored and the flows are assigned paths
according to their priority. B4 is also known for achieving near-optimal
network throughput. Microsoft's SWAN considers classes of priorities,
pertaining to critical, elastic and background traffic~\cite{Hong.2013}.
Network paths are first assigned per priority class. Within each coarse
assignment, a max-min fairness approach is used to distribute resources
to specific flows. Bell Labs propose a more direct approach, seeking
to solve the formal maximal link utilization problem, given explicit
flow requests~\cite{Agarwal.2013}. Other studies focus on scenarios
such as partially SDN-controlled networks, or on multipath routing~\cite{Domza.2015},
exploiting the monitoring capabilities of OpenFlow \cite{Domza.2015}.}

\textcolor{black}{Related studies have also demonstrated the SDN potential
in congestion handling, network monitoring and application specific
QoS. MicroTE~\cite{Benson.2011b}, Hedera~\cite{AlFares.2010} and
Mahout~\cite{Curtis.2011b} focus on the detection and special handling
of large \textquotedbl{}elephant\textquotedbl{} flows, under the assumption
that they constitute the usual suspects of congestion. Such flows
are assigned to paths which do not conflict with the bulk of the remaining
traffic. The network monitoring is continuous, scanning network-wide
for large flows via periodic polling at the scale of $5\,\mbox{sec}$.
A similar high-level logic is adopted for application-specific TE.
Initially, applications state their latency and bandwidth requirements.
Then, the corresponding flows are mapped to appropriate paths. PlugNserve
assumes application-specific TE and focuses on robustness against
adding/removing nodes in real-time \cite{Handigol.2009}. The Aster{*}x
approach seeks to keep the response time of web services as low as
possible, by monitoring the congestion levels of the network and picking
the less loaded routes per new flow \cite{Handigol.}. Authors in
\cite{Gvozdiev.2014} propose a scheme that assigns flows to routes,
selected from a list of paths with similar bandwidth, ordered by ascending
latency.}

\textcolor{black}{The scalability of SDN-enabled TE has constituted
an early consideration. Authors in \cite{McKeown.2008} stress the
need to reduce the control-plane load by: (i) minimizing the required
amount of flow rules installed to the network, and (ii) limiting the
interaction between the SDN controller and the routers. Towards the
first direction, Devoflow \cite{Mogul.2010} and DIFANE \cite{Yu.2010}
introduced an operation promoting wild-card rules, in order to minimize
the required flow entries. This fact is taken into consideration in
routing update processes, where older routing rules are retained for
some time, in order to transit smoothly from one TE instance to another
\cite{Reitblatt.2012,McGeer.2012,Katta.2013}. Towards the controller
workload direction, present solutions multiple controller deployments,
with workload balancing mechanisms \cite{Yu.2010,Hu.2012}. Kandoo,
for example, proposes a hierarchical controller deployment comprising
two layers of command. The lowest layer handles dedicated partitions
of the network, while a central controller coordinates any intra-partition
action \cite{HassasYeganeh.2012}.}

\textcolor{black}{Noticing the operational benefits and the growing
maturity of SDN, researchers studied its fitness as a platform for
evolving the inter-AS routing. The ossification of Internet routing,
the absence of QoS guarantees and the potentially slow convergence
and security issues of BGP constitute some of the long-standing issues~\cite{KontronisNetvolution}.
The 4D approach constituted an early }\textcolor{black}{\emph{clean-slate}}\textcolor{black}{{}
proposal for inter-AS routing~\cite{D4D}. 4D can be considered as
the precursor of OpenFlow, and allows AS operators to set clear network
objectives, obtain network-wide views and directly control the inter-network
state from a central point. Authors in~\cite{KontronisNetvolution}
proposed an evolvable platform for inter-AS routing based on the principle
of control-plane outsourcing. The routing logic of multiple ASes is
outsourced to an external trusted entity, which orchestrates the inter-AS
collaborative routing. SDN serves as the underlying technology, due
to the clear separation it enforces between the control and data planes.
Platforms that treat the routing infrastructure as a service constitute
specific, fitting choices (E.g., RouteFlow~\cite{routeflow}). In
order to limit the scale of the required changes in hardware and protocol
for adopting these solutions, studies on partial deployment have taken
place~\cite{PavlosMAMA2016}. Highly attractive gains in routing
convergence are attained when $50$\% of the ASes convert to the new
scheme. The SDX approach proposed the software-ization of Internet
Exchanged Points (IXPs), central rendezvous points for AS peering
that grow in popularity~\cite{SDX.2015}. The Control Exchange Points
approach proposes that ASes publish some internal paths to an external
entity~\cite{kotronisCXP}. The entity can then stitch together paths
crossing multiple ASes, offering end-to-end QoS upon demand. }

\textcolor{black}{The present work is based on the premise that ASes
may require a minimal-commitment scheme at first, in order to try-out
and evaluate the prospects of collaboration. Therefore, we propose
a TE approach that can operate over existing hardware and NCSes, without
introducing point-of-failure considerations. The proposed scheme constitutes
a novel application of BPR routing \cite{Georgiadis.2005}, and it
brings its analytically-proven, latency-aware throughput-optimality
to inter-AS TE, with limited commitment. The proposed approach prioritizes
compatibility with existing AS operations, differing from clean-slate
approaches~\cite{Ying.2008}. The proposed system is not intended
to offer the rich set of features promised by related studies, but
rather to serve as an intermediate step towards convincing ASes to
gradually adopt them. The scheme assumes SDN-inspired principles for
its operation, seeking to lay the foundations for more extensive SDN
inter-AS adoption in the future. ASes that adopt the proposed scheme
can benefit from the optimal stability it entails. CXP pathlet stitching
can serve as the next step towards closer collaboration, taking advantage
of the capabilities offered by SDX points for end-to-end QoS. }

\textcolor{black}{An early version of the proposed scheme is given
at~\cite{Liaskos.ISCC15}. \vspace{-5bp}
}

\section{\textcolor{black}{Conclusion\label{sec:Conclusion-and-Future}}}

\textcolor{black}{The present study proposed the application of BPR
routing as a minimal-commitment scheme for collaborative inter-AS
traffic engineering. BPR itself promises network throughput maximization
and transit traffic increase, which can supply an economic incentive
for AS participation to the proposed scheme. Novel BPR-based routing
algorithms where analyzed for the inter-AS traffic engineering scenario.
From a systemic aspect, BPR is deployed to the network of collaborating
ASes via a simple SDN-inspired interface, in the form of temporary
priority routing rules. Extensive simulations evaluated the further
advantages of the proposed scheme, namely stability under increased
network load and co-existence with existing inter-AS routing mechanisms.
Serving as a first step towards deeper AS cooperation could constitute
a significant application of the proposed approach.\vspace{-10bp}
}

\bibliographystyle{IEEEtran}

\begin{IEEEbiography}[{\crule[white]{0.001cm}{0.001cm} }]{}
\end{IEEEbiography}

\begin{IEEEbiography}[{\includegraphics[width=1in,height=1.25in]{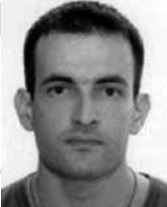}}]{Christos Liaskos}
 received the Diploma in Electrical and Computer Engineering from
the Aristotle University of Thessaloniki (AUTH), Greece in 2004, the
MSc degree in Medical Informatics in 2008 from the Medical School,
AUTH and the Ph.D. degree in Computer Networking from the Dept. of
Informatics, AUTH in 2014. He has published work in venues such as
IEEE Transactions on: Networking, Computers, Vehicular Technology,
Broadcasting, Systems Man \& Cybernetics, Communications, IEEE INFOCOM,
Elsevier NANOCOMNET, ACM NANOCOM, at a total of more than 40 publications,
and has received a best paper award. He is currently a researcher
at the Foundation of Research and Technology, Hellas (FORTH). His
research interest include traffic engineering and security in software-defined
networks, wireless networks and nanonetworks.
\end{IEEEbiography}

\begin{IEEEbiography}[{\includegraphics[width=1in,height=1.25in]{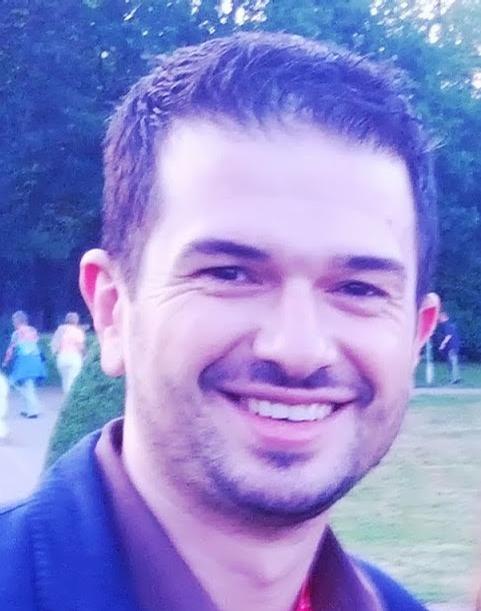}}]{Xenofontas Dimitropoulos}
 is an Assistant Professor at the Institute of Computer Science
Department, University of Crete (UoC) since January 2014. In addition,
he is affiliated with the Institute of Computer Science of the Foundation
for Research and Technology Hellas (FORTH). He leads a research group
working on Internet measurements and software defined networks with
the main goal of making the Internet more reliable and secure. Before,
he worked in the Georgia Institute of Technology, the IBM Research
Labs and the University of California San Diego (UCSD). He has published
more than 88 papers and 3 patents and has received 2 best paper awards.
In addition, he has received prestigious grants from the European
Research Council, the Marie Curie and the Fulbright programs. He has
served in the Organizing and Technical Program Committee of the flagship
networking conference, ACM SIGCOMM.
\end{IEEEbiography}

\begin{IEEEbiography}[{\includegraphics[width=1in,height=1.25in]{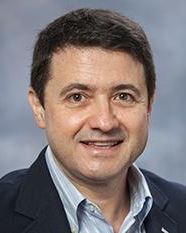}}]{Leandros Tassiulas}
 (S\textquoteright 89\textendash M\textquoteright 91\textendash SM\textquoteright 05\textendash F\textquoteright 07)
received the Ph.D. degree in electrical engineering from the University
of Maryland, College Park, MD, USA, in 1991. He has been a Faculty
Member with the NYU Polytechnic School of Engineering, Brooklyn, NY,
USA, the University of Maryland, College Park, and the University
of Thessaly, Volos, Greece. He is currently the John C. Malone Professor
of Electrical Engineering with Yale University, New Haven, CT, USA.
His most notable contributions include the max-weight scheduling algorithm
and the back-pressure network control policy, opportunistic scheduling
in wireless, the maximum lifetime approach for wireless network energy
management, and the consideration of joint access control and antenna
transmission management in multiple antenna wireless systems. His
research interests include computer and communication networks with
an emphasis on fundamental mathematical models and algorithms of complex
networks, architectures and protocols of wireless systems, sensor
networks, novel internet architectures, and experimental platforms
for network research. He was a recipient of several awards, including
the IEEE Koji Kobayashi Computer and Communications Award, the Inaugural
INFOCOM 2007 Achievement Award for fundamental contributions to resource
allocation in communication networks, the INFOCOM 1994 Best Paper
Award, the National Science Foundation (NSF) Research Initiation Award
(1992), the NSF CAREER Award (1995), the Office of Naval Research
Young Investigator Award (1997), and the Bodossaki Foundation Award
(1999).
\end{IEEEbiography}

\end{document}